\documentclass[reprint,aps,prd,nofootinbib, floatfix]{revtex4-1}
\usepackage[utf8]{inputenc}
\usepackage[utf8]{inputenc}
\usepackage{amsmath}
\usepackage{amsfonts}
\usepackage{amssymb}
\usepackage[left=2cm,right=2cm,top=2cm,bottom=2cm]{geometry}
\usepackage{braket}
\usepackage{dsfont}
\usepackage{hyperref}

\usepackage{graphics}
\usepackage{tikz}
\usetikzlibrary{arrows,decorations.pathmorphing,backgrounds,positioning,fit,petri,shapes,snakes}
\usepackage{lipsum}

\usepackage{multirow}
\begin{document}

\title{Preparing for the Cosmic Shear Data Flood: Optimal Data Extraction and Simulation Requirements for Stage IV Dark Energy Experiments}

\begin{abstract}
Upcoming photometric lensing surveys will considerably tighten constraints on the neutrino mass and the dark energy equation of state. 
Nevertheless it remains an open question of how to optimally extract the information and how well the matter power spectrum must be known to obtain unbiased cosmological parameter estimates.
By performing a Principal Component Analysis (PCA), we quantify the sensitivity of 3D cosmic shear and tomography with different binning strategies to different regions of the lensing kernel and matter power spectrum, and hence the background geometry and growth of structure in the Universe.
We find that a large number of equally spaced tomographic bins in redshift can extract nearly all the cosmological information without the additional computational expense of 3D cosmic shear.
Meanwhile a large fraction of the information comes from small poorly understood scales in the matter power spectrum, that can lead to biases on measurements of cosmological parameters.
In light of this, we define and compute a cosmology-independent measure of the bias due to imperfect knowledge of the power spectrum.
For a Euclid-like survey, we find that the power spectrum must be known to an accuracy of less than $1\%$ on scales with $k \leq 7 h \text{ Mpc} ^ {-1}.$
This requirement is not absolute since the bias depends on the magnitude of modelling errors, where they occur in $k$-$z$ space, and the correlation between them, all of which are specific to any particular model.
We therefore compute the bias in several of the most likely modelling scenarios and introduce a general formalism and public code, {\tt RequiSim}, to compute the expected bias from any non-linear model. 
\end{abstract}

\author{Peter L.~Taylor}
\email{peterllewelyntaylor@gmail.com}
\author{Thomas D.~Kitching}
\author{Jason D.~McEwen}
\affiliation{Mullard Space Science Laboratory, University College London, Holmbury St.~Mary, Dorking, Surrey RH5 6NT, UK}
\date{19 February 2018}
\maketitle

\section{Introduction}
As photons travel from distant galaxies their paths are gravitationally distorted by inhomogeneities in the gravitational field. 
This changes the ellipticities of observed galaxies, which on the largest scales is referred to as cosmic shear.
Since this shear signal is sensitive to both the growth of structure and the background geometry of the Universe, by measuring the statistics of these distortions over a large number of galaxies it is possible to constrain cosmological parameters \cite{refregier2004weak, heymans2013cfhtlens, hildebrandt2017kids,  troxel2017dark}.
\par As the number of source galaxies in most surveys peaks in the relatively low redshift Universe, below $z=2$, cosmic shear experiments are primarily sensitive to physics encoded by cosmological parameters that affect the late Universe. 
This makes weak lensing an ideal probe to distinguish between models of dark energy and determine the sum of the masses of neutrinos \cite{refregier2010euclid}.
\par We are entering a golden age of Stage IV weak lensing experiments \cite{albrecht2006report} as data from Euclid\footnote{\url{http://euclid-ec.org}} \cite{laureijs2010euclid}, WFIRST\footnote{\url{https://www.nasa.gov/wfirst}} \cite{spergel2015wide} and LSST\footnote{\url{https://www.lsst.org}}~\cite{anthony4836large} will be available within the next decade. Before these data sets arrive, it is important to determine the optimal method to extract cosmological information. 
\par At present there are two proposed methods that use shear information in different ways: so-called `tomography' and `3D cosmic shear', proposed in \cite{hushotnoise} and \cite{heavens3d} respectively. The former refers to identically weighting galaxies at all redshifts and compresses the data into \textit{tomographic bins} where all galaxies inside a certain redshift range are assigned the same redshift; using this approach, the expected errors on the dark energy equation of state parameters converges for approximately $10$-$20$ bins \cite{bridle2007dark}. Meanwhile the latter technique refers to a weighting scheme only, where data along the line of sight is given the spherical-Bessel weight that depends on radial and angular wave-numbers within a single redshift bin. We discuss the motivation for this weight in Appendix~\ref{sec:A}.
\par Comparing these two techniques is the first of objective of this paper. We compare tomography and 3D cosmic shear using the new Generalised Lensing and Shear Spectra ({\tt GLaSS}) code, soon to be made publicly available~\cite{taylor2018testing} as part of the {\tt Cosmosis} \cite{cosmosis} modular cosmological software package.
\par As no data compression takes place, it has been suggested that spherical-Bessel weighted lensing is more sensitive to radial information \cite{heavens3d, kitching20143d, spuriomancini}. However, in tomography, compressing the data by coarsely binning in the radial direction is not strictly necessary. Whilst it is true that for a very large number of redshift bins, shot noise will dominate in the intra-bin power spectra (`auto-correlation') of bins, the inter-bin power spectra (`cross-corrrelation') between bins is free of shot-noise, and will contain the majority of the information. We investigate these issues in this paper. 
\par In any case, Stage IV experiments offer at least an order of magnitude improvement in \textit{precision} over existing surveys. With increasing statistical precision it is important to keep systematics in check to ensure that measurements remain \textit{unbiased}, to avoid far reaching but incorrect conclusions. Potential sources of bias include photometric redshift errors \cite{efstathiou2017problems}, inaccurate intrinsic alignment models \cite{troxel2015intrinsic}, and instrumental inaccuracies and uncertainties \cite{massey2012origins}. 
\par An additional uncertainty comes from the difficultly in modelling non-linearities~\cite{takahashi2012revising} and baryonic effects \cite{rudd2008effects} at small scales (high $k$-modes), leading to inaccurate matter power spectrum models. There is no easy way to separate out the signal contributions from these small scales because small scale perturbations in the matter power spectrum at low-redshift contribute to the same modes in the signal as larger scale perturbations at higher redshift. To avoid contamination from these poorly understood scales, it is imperative to understand how bias in power spectrum modelling propagates through to bias on the lensing signal itself. This is the second objective of this work.
\par  In Section~\ref{Formalism} we derive expressions for the signal and the noise for the most general weighted two-point statistic and show how tomography and 3D cosmic shear are both just special cases. Next we present an heuristic guide to the Principal Component Analysis (PCA) formalism that we will use to assess the information content of these statistics. A more detailed exposition can be found in the Appendix~\ref{sec:PCA}. We also show how biases in the matter power spectrum modelling can be summarised in terms of a `knowledge matrix' (that encodes the level of knowledge one has about the matter power spectrum model, including correlations between $k$-modes and redshift) and present a cosmological parameter-independent measure of the bias in the lensing signal due to matter power spectrum modelling errors. In Section~\ref{Results}, we determine how 3D cosmic shear and tomographic lensing are sensitive to the matter power spectrum and the lensing kernel, using the formalism presented in Section~\ref{Formalism}. We show the bias in the lensing statistics induced by biases in the matter power spectrum for a variety of knowledge matrices corresponding to realistic cases.  
\par Our code, {\tt RequiSim}, used in this last part is made publicly available and can be used to assess whether any given matter power spectrum simulation or model is accurate enough for a Euclid-like lensing survey. Finally in the Appendices we outline our modelling choices, discuss motivations for the spherical-bessel weight in 3D cosmic shear, provide details about PCA formalism, address the challenges of removing sensitivity to small scales in the matter power spectrum and provide information about {\tt RequiSim}.

    \section{Cosmic Shear Formalism} \label{Formalism}
    \subsection{The Generalised Spherical-Transform}
The shear field is defined everywhere, but it can only be sampled at the position of galaxies. We can transform the sampled shear field into the spherical-Bessel basis. This is commonly referred to as `3D cosmic shear', or `3D weak lensing'. In this case the shear field is given by:
\begin{equation}
\label{sB}
\gamma_{\ell m} \left( \eta \right) = \sqrt{\frac{2}{\pi}} \sum _g \gamma_g \left( r_g, \boldsymbol{\theta_g } \right)  j_{\ell} \left( \eta r_g \right) {}_2 Y_{\ell m} \left( \boldsymbol {\theta_g} \right),
\end{equation}
where the sum is over all galaxies $g$, $ j_{\ell} \left( \eta r_g \right)$ are spherical Bessel functions and $ _2 Y_{\ell m} \left( \boldsymbol {\theta_g} \right)$ are spin-2 spherical harmonics. Motivations for the Bessel weight are discussed in the Appendix~\ref{sec:A}. Here we explicitly write the harmonic variable as $\eta$, so that it is not confused with $k$, which is used to denote the wave-number in the matter power spectrum only.

\par As discussed in \cite{heavens2006measuring}, we could also weight the data by an arbitrary weight function, $W$. In \cite{heavens2006measuring} this is taken to be a function of the co-moving distance $r$ only, but in general the weight function can also depend on a radial and angular wave-number $\eta$ and $\ell$: $W_{\ell}\left(\eta, r \right)$. Weights were also considered in \cite{ayaita2012investigating, schafer2012weak}, but we consider a more general formalism. Replacing the Bessel functions with a general weight, we define \textit{the generalised spherical-transform} given by: 
\begin{equation} \label{eq:transform}
\gamma_{\ell m} \left( \eta \right) =  \sqrt{\frac{2}{\pi}} \sum _g \gamma_g \left( r_g, \boldsymbol{\theta_g } \right)W_{\ell} \left(\eta, r_g \right) {}_2 Y_{\ell m} \left( \boldsymbol {\theta_g} \right), 
\end{equation}
where $\eta$ is a label that can be a wave-number or a real-space quantity. The expression for the lensing matter power spectrum becomes:
\begin{equation} \label{eq:c_l}
\begin{aligned}
C_{\ell}^{\gamma \gamma} \left( \eta_1, \eta_2 \right) =  \frac{9 \Omega_m ^ 2 H_0 ^ 4}{16 \pi^4 c^ 4 }\frac{\left( \ell + 2 \right)!}{\left( \ell - 2 \right)!} & \int \frac{\text{d} k}{ {k} ^2} G_{\ell}^\gamma \left( \eta_1, k \right)\\ & \times G_{\ell}^\gamma \left(\eta_2, k \right) ,
\end{aligned}
\end{equation}
where $\Omega_m$ is the fractional energy density of matter, $c$ is the speed of light in a vacuum and $H_0$ is the present day Hubble constant. The G-matrix is defined as:
\begin{equation} \label{eq:G}
\begin{aligned}
G_{\ell} ^ \gamma \left( \eta , k \right) \equiv \int \text{d}z_p \text{d} z' \text{ }  &n \left(z_p \right) p \left(z' | z_p \right) \\ & \times W_{\ell} \left(\eta, r  \left[z' \right]\right) U_{\ell} \left(r \left[ z' \right], k \right)  
\end{aligned}
\end{equation}
and the $U$-matrix, which contains all the cosmological information, is given by:
\begin{equation} \label{eq:U}
U_{\ell} \left(r[z], k \right) \equiv \int ^ r _0 \text{d} r' \text{ } \frac{F_K \left(r, r' \right)}{a \left(r' \right)} j_{\ell} \left( k r' \right) P^{1/2 }\left(k ; r' \right).
\end{equation}
In the above expressions $n(z)$ is the radial distribution of galaxies and photometric uncertainty $p \left( z|z' \right)$ gives the probability that a galaxy has a redshift $z$, given a photometric redshift measurement $z'$, $P \left(k,r \right)$ is the matter power spectrum, $r$ is the co-moving distance and the lensing kernel, $F_K \left(r, r' \right)$, is defined as:
\begin{equation} \label{eq:lens_kernel}
 F_K \left(r, r' \right)\equiv \frac{r-r'}{rr'}, 
\end{equation}
for a flat cosmology. We are implicitly assuming equal-time power spectrum throughout. This has been found to be a good approximation \cite{kitching2017unequal}.
\par The weights also propagate to the shot noise \cite{heavens2006measuring}, which becomes:
\begin{equation} \label{eq:Noise}
N_\ell^{ e e} \left( \eta_1, \eta_2 \right) = \frac{\sigma_e ^2}{2 \pi ^ 2} \int \text{d} z \text{ } n\left( z \right)W_\ell \left(\eta_1, r \right)W_\ell \left(\eta_2, r \right) , 
\end{equation}
where $\sigma_e ^2$ is the variance of the intrinsic ellipticities in galaxies. We take $\sigma_e  = 0.3$ throughout \cite{brown2003shear}.
\par When taking the weight-function, $W_\ell \left(\eta, r  \left( z \right)  \right) = j_\ell \left(\eta r\right)$ in equations~\ref{eq:G} and~\ref{eq:Noise}, the normal 3D cosmic shear equations are recovered. For the bin associated with redshift region $I$, in the tomographic case, we take the weight function, $W^I$, in equation~\ref{eq:G} as a top hat function in redshift only, so that:
\begin{equation}
   W ^ I \left(z \right) \equiv
    \begin{cases}
      1 & \text{if $z \in I$  }\\
      0 & \text{if $z \notin I$  }\\
    \end{cases} 
\end{equation}
and the shot noise reduces to:
\begin{equation}
N_{ij} = \frac{\sigma_ e ^ 2}{2 \pi ^2} N_i \delta_{ij},
\end{equation}
where $i$ and $j$ label the bin numbers and $N_i$ is the number of galaxies in bin $i$.
\par The expressions above can be simplified further. As found in~\cite{kitchinglimits}, the extended Limber approximation, given in~\cite{loverdelimber}, is sufficiently accurate for $\ell > 100$. Then the $U$-matrix becomes:
\begin{equation} \label{eq:limber}
U_{\ell} \left(r, k \right) = \frac{F_k \left( r, \nu \left( k \right) \right)}{k a \left( \nu\left( k \right)  \right)} \sqrt {\frac{\pi}{2 \left( {\ell} + 1/2 \right)}}  P ^ {1/2} \left( k, \nu\left( k \right)  \right),
\end{equation}
where $\nu\left( k \right)  \equiv \frac{{\ell}+ 1/2}{k}$. This implies that for fixed $\ell$-modes above  $\ell\approx 100$, the signal is sensitive to the power spectrum only along the curve $\ell+ 1/2 = kr$ \cite{loverdelimber}. We refer to these curves as \textit{Limber lines} in the $(k$,$r[z])$ plane (they are plotted in Figure \ref{fig:tomo_fig_high_res} for different $\ell$-modes).
\par The pre-factors used in the equations above for the signal and noise do not follow the standard notation. However once we have made the Limber approximation (given in equation~\ref{eq:limber}), our expressions reduce to the usual ones given in~\cite{hushotnoise, joachimi2010simultaneous}. Our convention is taken to be consistent with the one used in \cite{heavens3d}.

\subsection{A Review of Cosmic Shear Likelihood Analyses} \label{sec:likelihood}
We review the details of the likelihood analysis used to extract cosmological information from cosmic shear data. This has been used successfully to constrain cosmological parameters in \cite{heymans2013cfhtlens, troxel2017dark, hildebrandt2016kids}. 
\par Assuming the likelihood is Gaussian, the likelihood for a set of parameters, $p$, is:
\begin{equation}
\text{ln } \mathcal{L}\left(p \right) = - \frac{1}{2} \sum_{a,b} \left[ D_a - T_a\left( p \right) \right] C_{ab}^{-1} \left[ D_b - T_b \left( p\right) \right],
\end{equation}
where $D_a$ is a data vector, $T_a$ is a theory vector and $C_{ab}^{-1}$ is the inverse covariance matrix between elements of the data vector. 
\par The covariance can be computed from ray-tracing simulations~\cite{heymans2013cfhtlens}, bootstrapped directly from the data~\cite{massey2007cosmos}, computed from fast lognormal simulations~\cite{troxel2017dark,xavier2016improving,mancini20183d} or by approximating the shear field as a Gaussian random field~\cite{joachimi2008analysis}. A thorough discussion of the relative merits of each of these approaches is given in~\cite{heymans2013cfhtlens}. 
\par In a power spectrum analysis, the data vector is the spectrum of the observed generalised spherical-transformed shear field defined in equation (\ref{eq:transform}). Meanwhile the theory vector is computed from equation (\ref{eq:c_l}). 
\par In practice it is difficult to account for the impact of mask on the theory vector. For this reason, most analyses take the data and theory vectors as two-point correlation functions ${\xi}_{\pm}$. This data vector is readily computed from a shear catalogue and the theory vector is found by first computing the raw lensing spectrum and then applying a filter. More details can be found in~\cite{kitchinglimits}. The filter preferentially weights the sensitivity to certain modes in the lensing spectrum. We ignore this complication in this work and focus on the information contained in the raw lensing power spectrum. 
\par Once the covariance matrix is known we can sample to compute the posterior of the cosmological parameters. The computation time is dominated by the calculation of the lensing spectrum and in particular the computation of the matter power spectrum. It is infeasible to run a full N-body simulation at each point in parameter space so we must resort to fast emulator simulations to compute the matter power spectrum. These suffer from a loss of accuracy at small scales.
\par In the remainder of this paper we try to answer the following two questions:
\begin{itemize}
\item{Assuming that the covariance matrix can be computed to sufficient accuracy, what is the optimal transform weight in equation (\ref{eq:transform})? If we further restrict our attention just to the tomographic case, what is the optimal binning strategy?}
\item{Since the theory vector depends on the matter power spectrum, to what accuracy does this have to be known to obtain an unbiased cosmological measurement?}
\end{itemize}

\subsection{Tomographic Binning Strategy} \label{sec:binning}

   \begin{figure}
   \centering
    \vspace{2mm}
    \includegraphics[width=85mm]{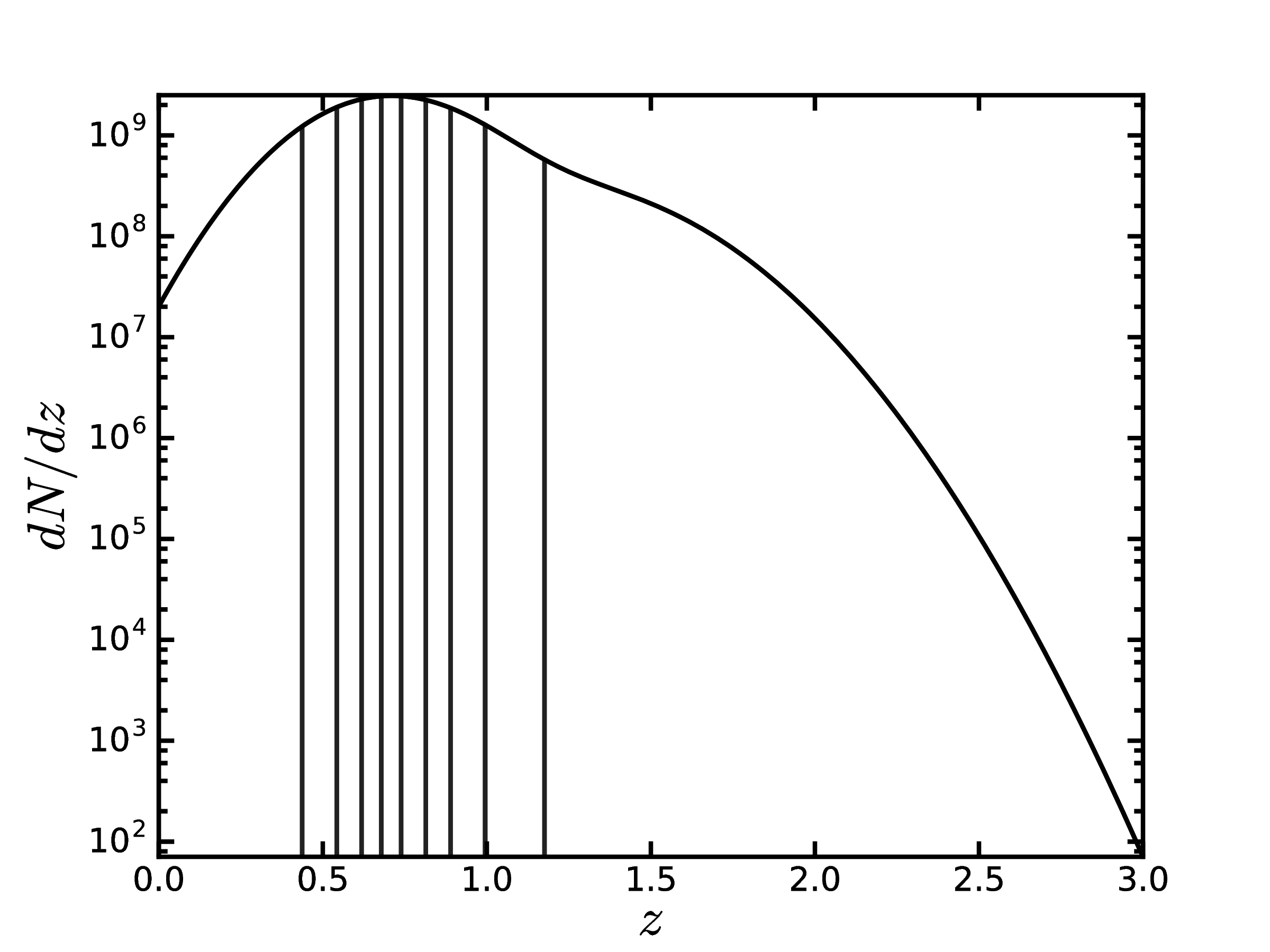}
    \includegraphics[width=85mm]{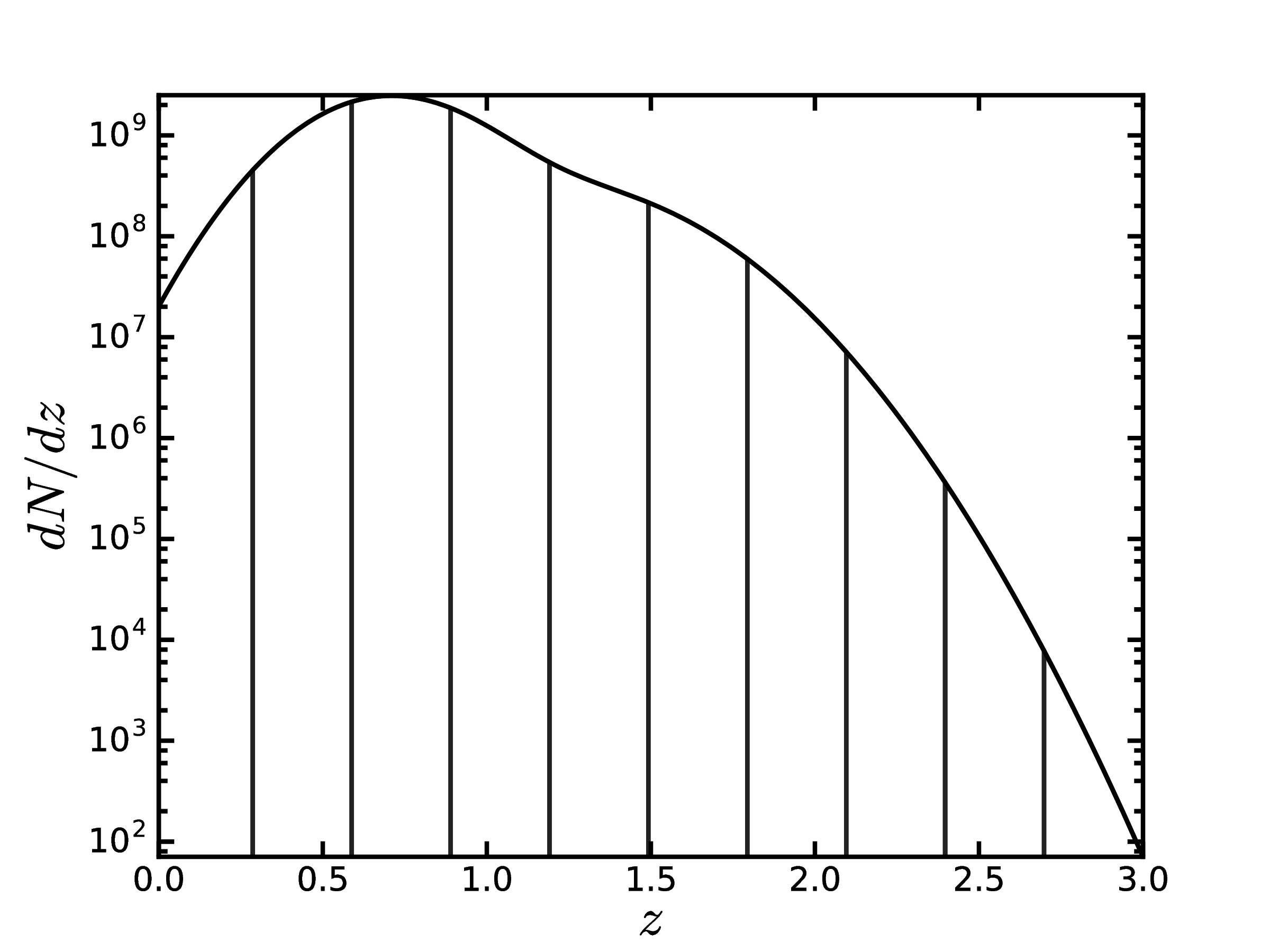}
    \caption{Differential number of galaxies as a function of redshift for two different tomographic binning strategies. {\bf top}: $10$-bins with an equal number of galaxies per bin. Equal galaxy binning captures most of the information with a small number of bins (see Figure \ref{fig:bin_convergence}), but an extremely large number of bins would be needed to capture the high-redshift information since the highest redhsift bin is so wide. For a larger number of bins, near the peak of the distribution the bin width would fall below the resolution of the computation grid. {\bf bottom}: $10$-bins spaced equally in redshift. This binning strategy does not capture as much information for a small number of bins (see Figure \ref{fig:bin_convergence}), but it is easy to increase the number of bins.  }
    \label{fig:n_z_plot}
    \end{figure}

\par In this paper we investigate two different tomographic binning schemes: equal galaxies per bin and equally spaced redshift bins.  These are shown for $10$-bin tomography in Figure~\ref{fig:n_z_plot}. We have overlaid the bin boundaries onto the predicted Euclid wide-field differential galaxy distribution, defined by: $\frac{\partial N \left( z\right)}{\partial z} \equiv N_g \frac{\partial n \left( z\right)}{\partial z}$, where $N_g$ is the number of galaxies in the survey. An equal number of galaxies per bin is the convention; and for less than $20$ bins, it captures more information than equally spaced bins in redshift (see Figure~\ref{fig:bin_convergence} and the discussion in Sections \ref{lens kernel}-\ref{power spec}).  However to completely capture the 3D information around $100$-bins must be used (see Section \ref{lens kernel}-\ref{power spec}). This presents two challenges for the equal number of galaxies per bin scheme:
\begin{itemize}
\item{Since the distribution of galaxies, $n(z)$, falls sharply for high redshifts the last bin will span a large $z$-range. Thus high redshift information will be lost unless an incredibly large number of bins are used.} 
\item{Meanwhile if more than $20$ bins are used the width of the tomographic bins near the peak of $n(z)$ becomes very small. Placing an equal number of galaxies into each bin would require globally increasing the resolution, slowing the computation time.}
\end{itemize}
For the later reason, we will only consider tomography with an equal number of galaxies per bin, up to $20$-bins.
\par While it is unconventional to use over $20$ bins in a likelihood analysis, we stress that it barely increases the computation time of the theory vector. Using our algorithm, it takes $1.55$ seconds to compute the lensing spectrum once the matter power spectrum has been computed for $100$-bins, compared to $1.40$ seconds for $10$-bins on a single {\tt 2.7 GHz Intel i5 Core} on a 2015 Macbook Pro with $8$ GB of RAM. We stress that although it is easy to compute the theory vector with a large number of bins, it may be difficult to compute a significantly accurate inverse covariance due to numerical noise~\cite{heymans2013cfhtlens}.

\section{Principle Components and Bias} \label{sec:formalism2}

\subsection{An Heuristic Guide to Principle Component Analysis (PCA)} \label{sec:pca_guide}
Assuming a fiducial cosmology it is possible to compute the lensing spectra given in equation~\ref{eq:c_l}. Then for any set of parameters, $\{ \theta_i \}$, we can estimate the constraining power of a lensing survey using the Fisher matrix formalism. Details are given in the Appendix~\ref{sec:B}.
\par
\begin{figure*}
\centering
\includegraphics[width=0.30\linewidth]{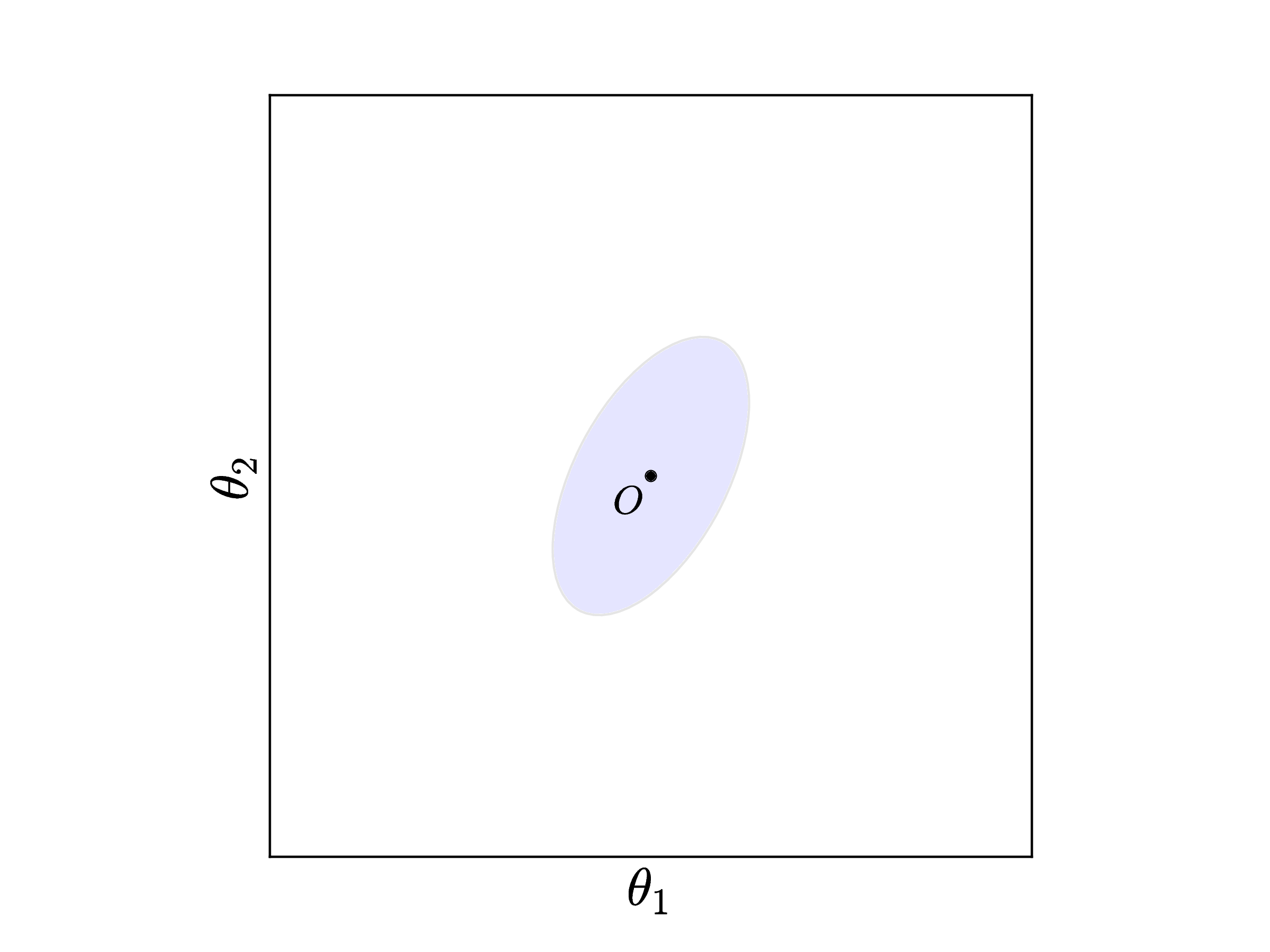}
\includegraphics[width=0.30\linewidth]{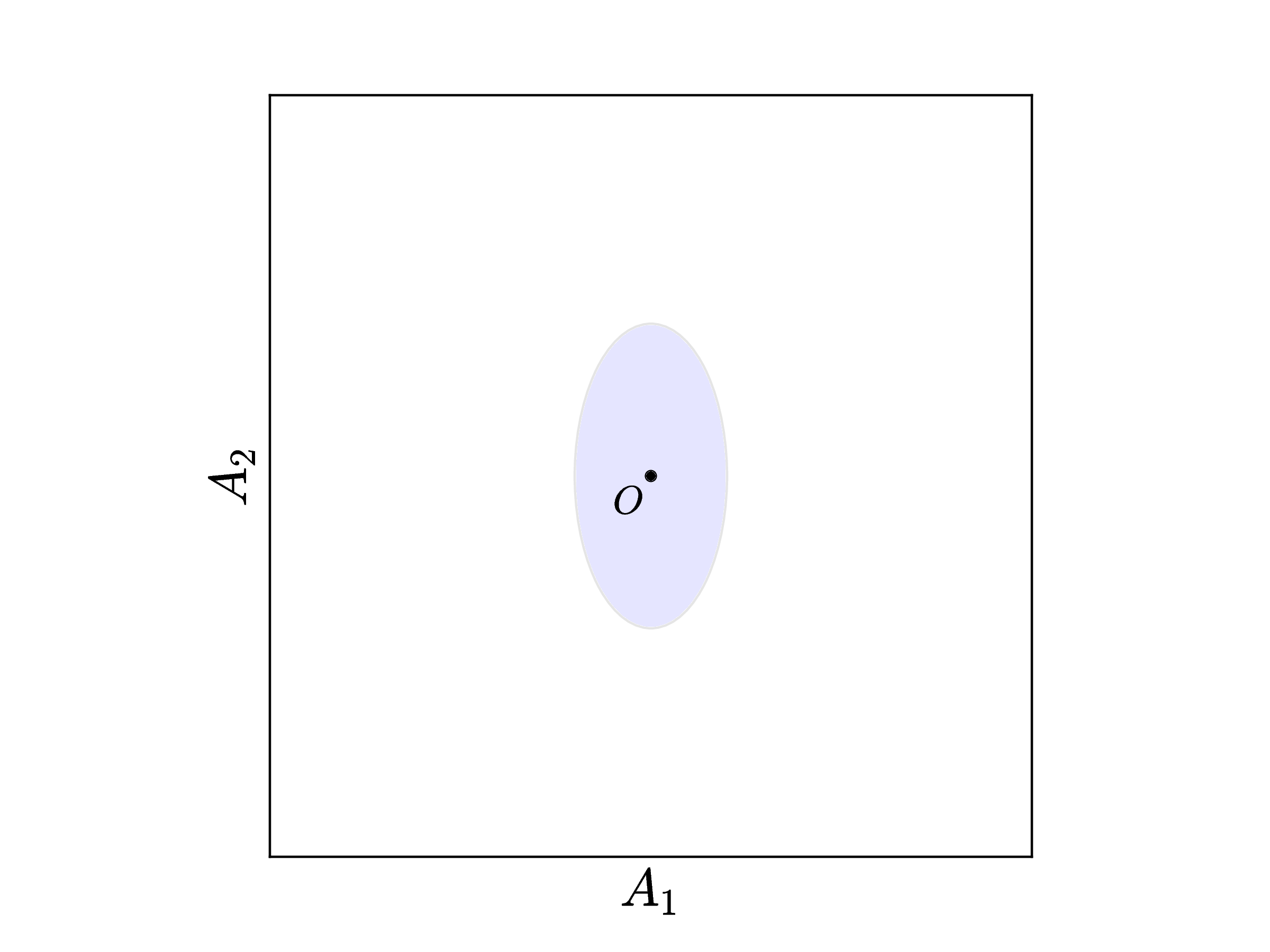}
\includegraphics[width=0.30\linewidth]{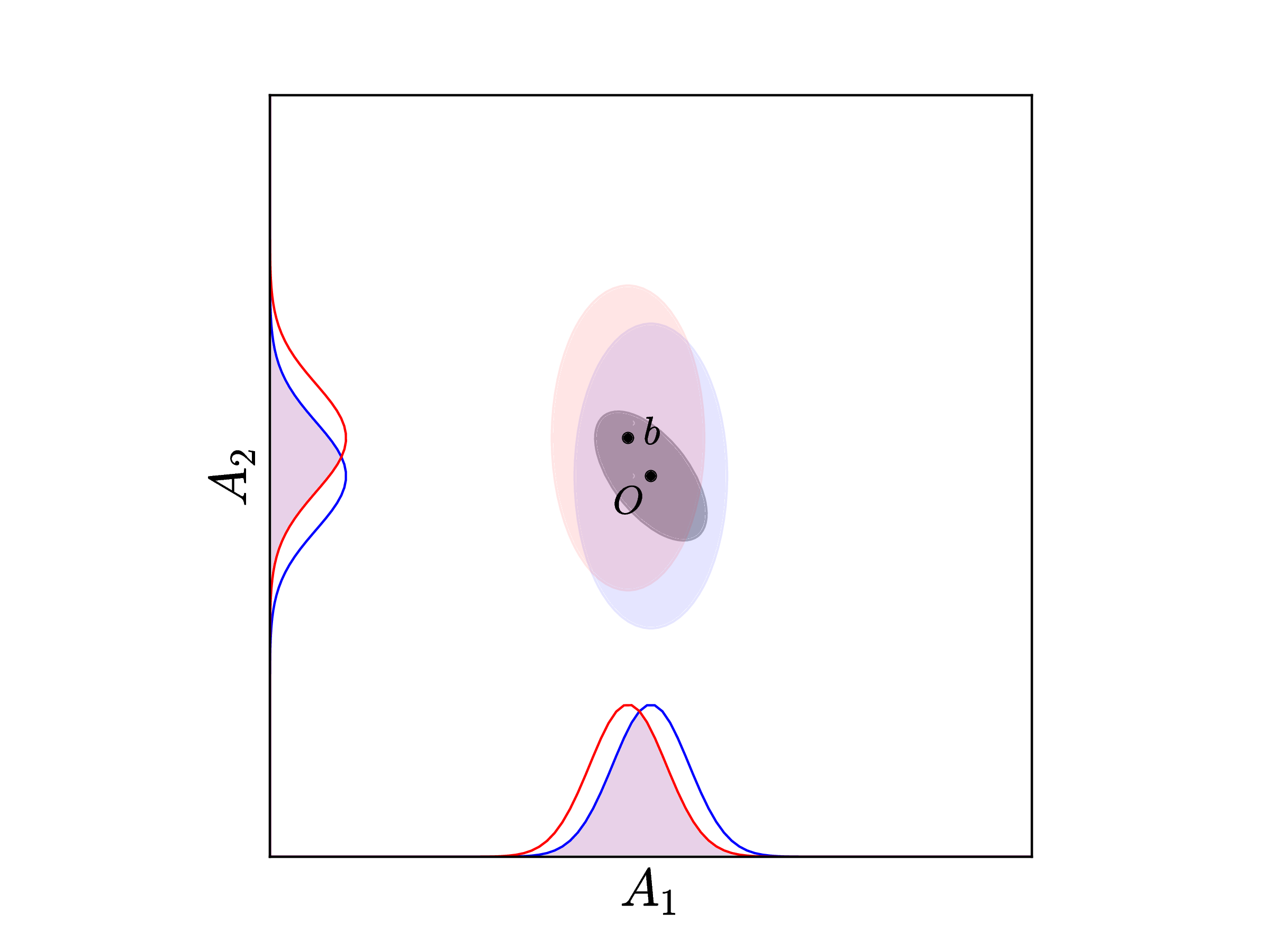}
\caption{This figure illustrates the procedure of calculating the principle components (see Section \ref{sec:pca_guide}) and the Variance Weighted Overlap (see Section \ref{sec:vwo}), which measures the bias in higher dimensional spaces {\bf left:} Parameter constraints for two degenerate parameters $\theta_1$ and $\theta_2$ from a cosmic shear experiment {\bf middle:} Rotation into a new basis of parameters that are independent. {\bf right:}  A bias, $b$, is drawn from the dark hashed oval representing the covariance on our modelling uncertainty. The new shifted covariance, centred on $b$, is represented by the red oval. The overlap integral between the shifted and unshifted distribution marginalised onto each direction is shown. Weighting these by the inverse variance on parameters $A_1$ and $A_2$ and marginalising over all shifts defines the Variance Weighted Overlap R($P_{\rm VWO}$).}
 \label{fig:heuristic}
\end{figure*}

 The parameters are often degenerate. For example, the purple oval in Figure \ref{fig:heuristic} represents the $2 \sigma $ confidence limit on parameters $\theta_1$ and $\theta_2$ estimated from a Fisher matrix analysis. Nevertheless it is possible to  rotate into a new basis of parameters, $ \{ A_i \}$, which are independent as in the second panel of Figure \ref{fig:heuristic}.
\par In this paper we will divide the co-moving distance $r(z)$ and the temporally evolving power spectrum $P(k,z)$ into cells in $z$ and $z$-$k$, respectively, closely following the analysis of \cite{pca}. We estimate constraints on the amplitude of these cells from cosmic shear using the Fisher matrix formalism. As in the example above, the amplitude of these cells are expected to be degenerate, so we find new linear combinations of these cells called components, whose amplitude are independent. Thus the components extract independent information. 
\par The components that are most tightly constrained are called the principle components (PCs). The sum of the inverse variance of each of the components which we write as, $I_{tot}$, called the \textit{total information content} is a figure of merit for the constraining power of an experiment. \par We investigate how many PCs are needed to extract the majority of the information. Ordering the components from the most to the least tightly constrained we compute \textit{Information Fraction} extracted as a function of the number of principle components.
\par Since we are interested in where information from the lensing kernel (equation ~\ref{eq:lens_kernel}) and power spectrum come from in $k$-$z$ space, we compute the \textit{Cramer-Rao Bound} on the amplitude inside each cell ($\sigma_{CRB} (r(z))$ and $\sigma_{CRB} (P(k,z)$ for co-moving distance and power spectrum cells respectively). We also compute the \textit{Variance Weighted Sum} ($S_{vw}$) on the amplitude of each cell. The first measures the inverse conditional error on each cell, while the second measures how tightly different regions of the co-moving distance and power spectrum are constrained, while accounting for correlations between cells. More details can be found in the Appendices~\ref{sec:PCA}-\ref{sec:D}.
\par Since the information about the co-moving distance is contained in the lensing kernel, we refer to the co-moving distance PCs as lensing kernel PCs for the the remainder of the text.

\subsection{Review of the Overlap Integral Formalism}

As well as understanding where lensing information originates, we must also understand the bias. The statistical error expected from upcoming surveys is fixed by the survey volume and the number of observed galaxies. To obtain unbiased results without increasing the statistical error, contributions from all sources of bias must be kept below a certain threshold.  We  review the formalism in~\cite{massey2012origins} (hereafter M12), which defines this threshold. 
\par M12 considered an experiment which measured a parameter with a Gaussian likelihood and statistical uncertainty $\sigma$. The bias, $b$, shifted the likelihood distribution, but did not change its shape. The distance between the two distributions was quantified by the overlap integral between the shifted and unshifted distributions: 
\begin{equation} \label{eq:overlap}
p_{\rm overlap} \left( b \right) = 1 -  \text{erf} \left( \frac{1}{2 \sqrt{2}} \frac{|b|}{\sigma}\right),
\end{equation}
If this was greater than $0.95$ (or less conservatively $0.90$), then the results are said to be \textit{unbiased}.
\par However, if the exact value of the bias was known it could be subtracted off, so the authors of M12 reinterpreted $b$ as a $95 \%$ confidence limit on the true bias. If the knowledge of the bias is also normally distributed, its standard deviation is then $\sigma_b = b /2$. Marginalising over $p_{\rm overlap} \left( b \right)$ by drawing $b$ values from this distribution and using the same overlap criteria as before, defines requirements on the magnitude of the bias $|b|$. These are: $|b| < 0.31 \sigma (0.62 \sigma) $ for a $95 \% (90 \%)$ overlap.
\par M12 used this formalism to place requirements on the total systematic bias needed for an unbiased measurement of the dark energy equation of state. However Stage IV cosmic shear experiments will place new constraints on other interesting parameters like the sum of neutrino masses and $w_a$ in the Chevalier-Polarski-Linder Parametrization \cite{chevallier2001accelerating,linder2003exploring}. We need to ensure that these, and parameters in any other cosmological parametrisation, will not be biased from inaccurate models of the power spectrum.
\par Perturbations to the lensing kernel and matter power spectrum will not have the same impact on the lensing spectrum. Thus we assume that these two varieties of PCs are only weakly correlated. Then, to check whether inaccurate power spectrum models induce bias, it is sufficient to ensure that power spectrum PCs are unbiased, and we can ignore bias propagating into the lensing kernel PCs. This is done by generalising the 1D overlap integral formalism to higher dimensions in the next section. 

\subsection{The Variance Weighted Overlap} \label{sec:vwo}
\par In analogy with the 1D case, we compute a higher-dimensional overlap integral between a biased and unbiased distribution. We envision a high dimensional parameter space of PC amplitudes and we will measure a generalisation of the overlap
between biased an unbiased probability distributions of these amplitudes. 
\par Since the parameters of interest are the power spectrum PCs, the distribution of biases in the unbiased case is taken to be the multivariate Gaussian with mean of zero and a covariance calculated from inverting the diagonal Fisher matrix in PCA-space. 
\par The distribution in the biased case has a shift in the mean (the multivariate equivalent of $|b|$) that is drawn from a Gaussian with mean zero and covariance $K$. We refer to this covariance as the \textit{knowledge matrix}, since it describes our confidence in our knowledge of the power spectrum. The $2 \sigma$ confidence region from this covariance is represented by the dark hashed ellipse in Figure~\ref{fig:heuristic}. We interpret $b_i$ as a $95 \%$ confidence limit on the true bias on principal component $i$ \footnote{In practice the knowledge matrix should be found in $k$-$z$ space, where it is known for a given simulation, and rotated into PCA-space. }. The elements of $K$ are then defined by:
\begin{equation}
K_{ij} \equiv \frac{b_i}{2}\frac{b_j}{2}.
\end{equation} 
The values of the resulting marginalised overlap integral are uninformative because they depend on the number of PCs included beyond those which contain $99 \%$ of the information (intuition about hyper-volumes in higher dimensional spaces is often wrong \cite{friedman2001elements}).
\par We instead define a new measure called the \textit{Variance Weighted Overlap} ($P_{\rm VWO}$), which reduces to the marginalised overlap integral in 1D used in \cite{massey2012origins}. We draw shifts in the mean from a multivariate Gaussian with the covariance given by the knowledge matrix. Then, instead of computing the hyper-volume of the overlap region, we marginalise to compute the 1D overlaps in each direction forming a set of 1D overlap volumes $\{p^{i}\}$ for each principal component $i$.
\par Since not all PCs are equally important we take an inverse variance weighted average over this set to form the Variance Weighted Probability Overlap given by:
\begin{equation}
P_{VWO} \equiv \frac{\sum_i w_i p ^i}{\sum_i w_i },
\end{equation}
where $w_i$ is the inverse variance on principal component $i$. Keeping consistency with M12, if $P_{\rm VWO} > 95 \%$ then the result is said to be `unbiased'. The requirements using this formalism are described in Section~\ref{sec:sim req}.
\par This procedure is illustrated in the right panel of Figure~\ref{fig:heuristic}. The blue oval represents the unshifted measured covariance centred on the origin $O$. A bias, $b$, is drawn from the dark hashed oval representing the covariance on our modelling uncertainty. The new shifted covariance, centred on $b$, is represented by the red oval. Marginalising onto each axis, we compute the overlap integral between the shifted and unshifted distribution on each parameter. After weighting these by the inverse variance on each component and marginalising over all shifts, we find the Variance Weighted Overlap ($P_{\rm VWO}$).

\section {Results} \label{Results}
	\subsection{Lensing Kernel PCs} \label{lens kernel}

\begin{table*}[t]
  \centering
    \caption{The total information content, $I_{tot}$, (see equation \ref{eq:info}) in the lensing kernel and the matter power spectrum. Specifically we compare: 3D cosmic shear, super tomography ($100$ equally space bins in redshift) and $10$-bin tomography in {\it GLaSS}. {\bf Case I}: (High resolution PCA grid and low resolution computation gird) Super tomography outperforms 10-bin tomography because there is no data compression in the former case. It also outperforms 3D cosmic shear due, but only due to slow convergence of the later technique (see Case II). {\bf Case II}: (Low resolution PCA grid and high resolution computation gird) Numerically, 3D cosmic shear has converged to within $3 \%$ of Super tomography. } 
  \begin{tabular}{l|c@{\hskip 14pt}c@{\hskip 14pt}c@{\hskip 14pt}|c@{\hskip 14pt}c}
  \hline \hline
    \multirow{2}{*}{{\bf PCA Run}} 
      & \multicolumn{3}{c|}{{\bf Case I}} 
          & \multicolumn{2}{c}{{\bf Case II}} \\    
   {\bf statistic} & super tomography & $\text{3D}$\textsuperscript{\ref{note1}} &  $10$-bin tomography & 3D\footnote{\label{note1}Due to the slow numerical of 3D cosmic shear, the total information content of is not fully converged for 3D lensing case (see Appendix~\ref{sec:convergence checks} and Figure~\ref{fig:convergence}).} & $10$-bin tomography \\
    \hline
    {\bf $I_{tot}$ (lensing kernel)}   & $1.00$ & $0.94$  & $0.94$ & $1.00$ & $0.97$  \\
    {\bf $I_{tot}$ (power spectrum)}   & $1.00$ & $0.77$  & $0.99$ & $1.00$ & $0.97$  \\
    \hline \hline
  \end{tabular}
  \label{tab:1}
\end{table*}

\begin{figure*}
	\begin{minipage}{1.0\textwidth}
		\includegraphics[width=5.5cm]{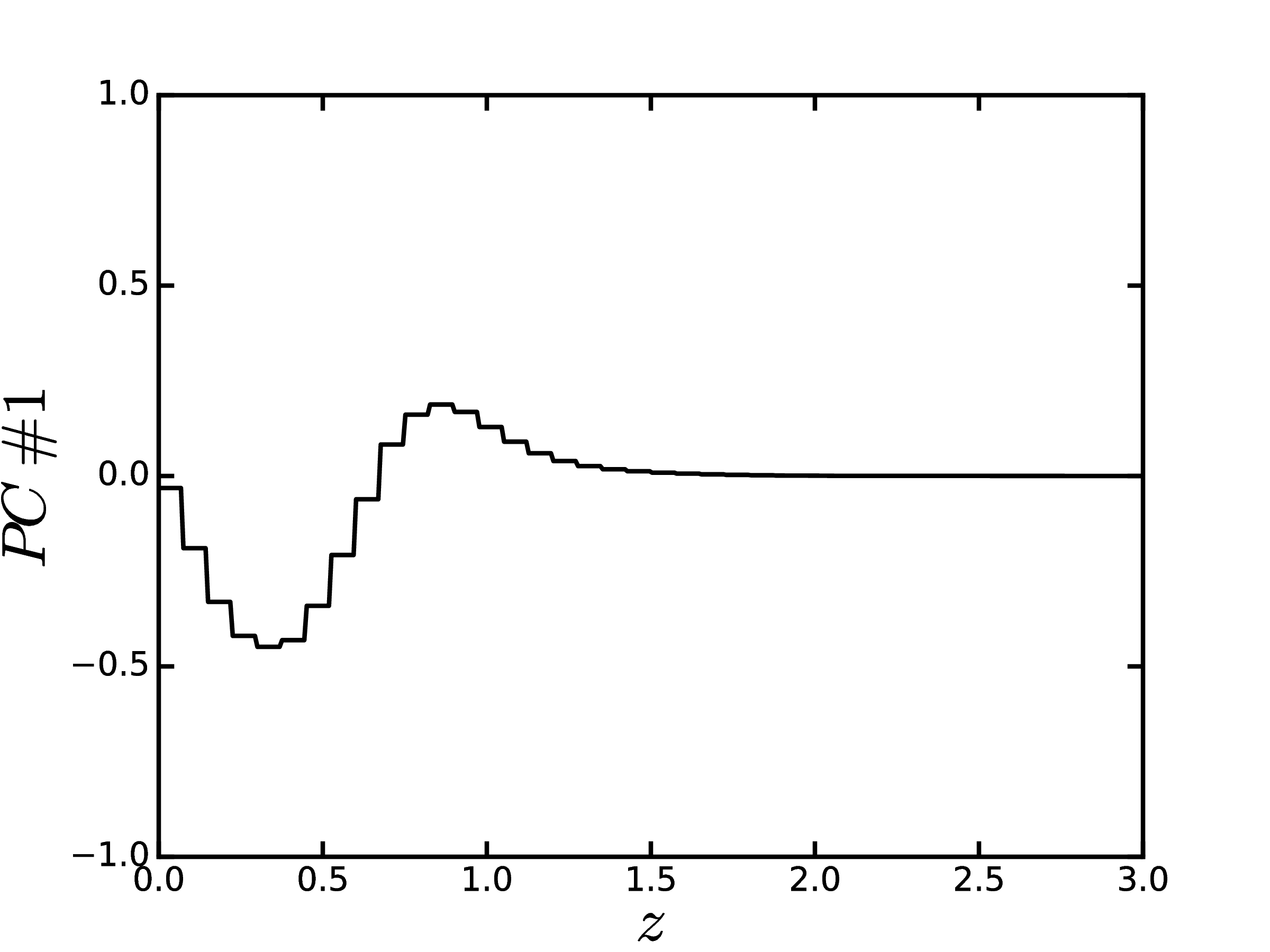}
		\includegraphics[width=5.5cm]{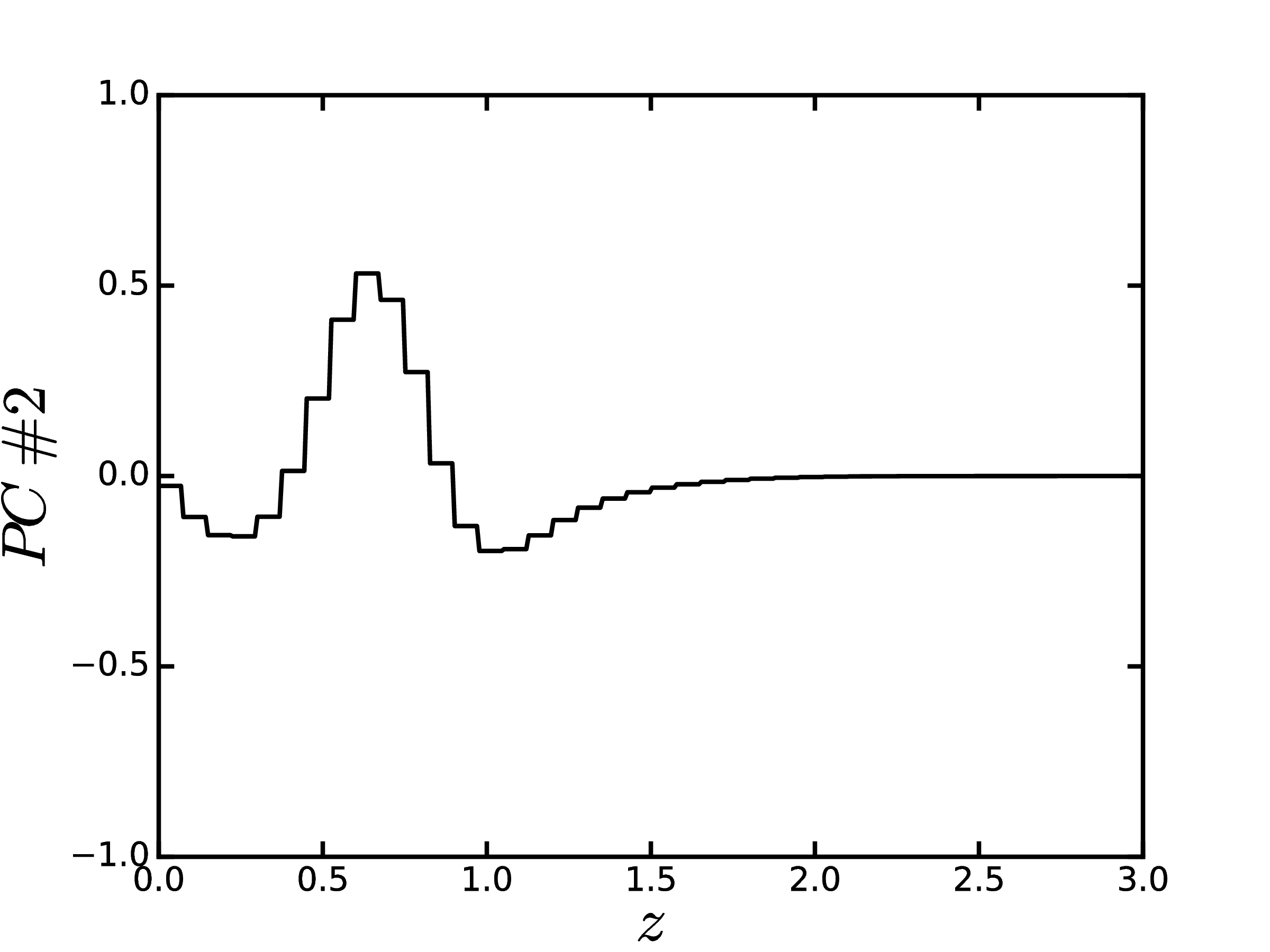}
		\includegraphics[width=5.5cm]{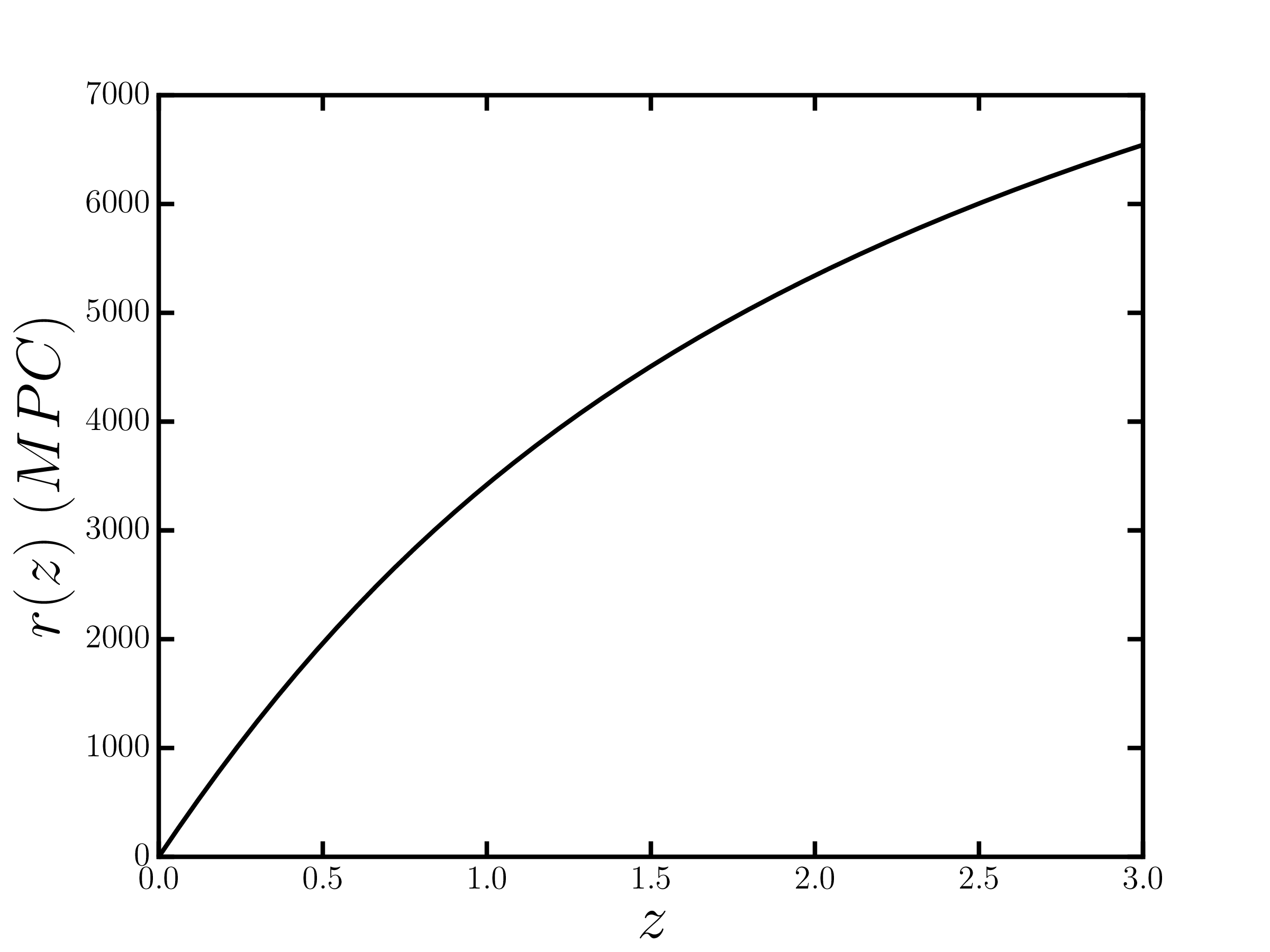}
        \includegraphics[width=5.5cm]{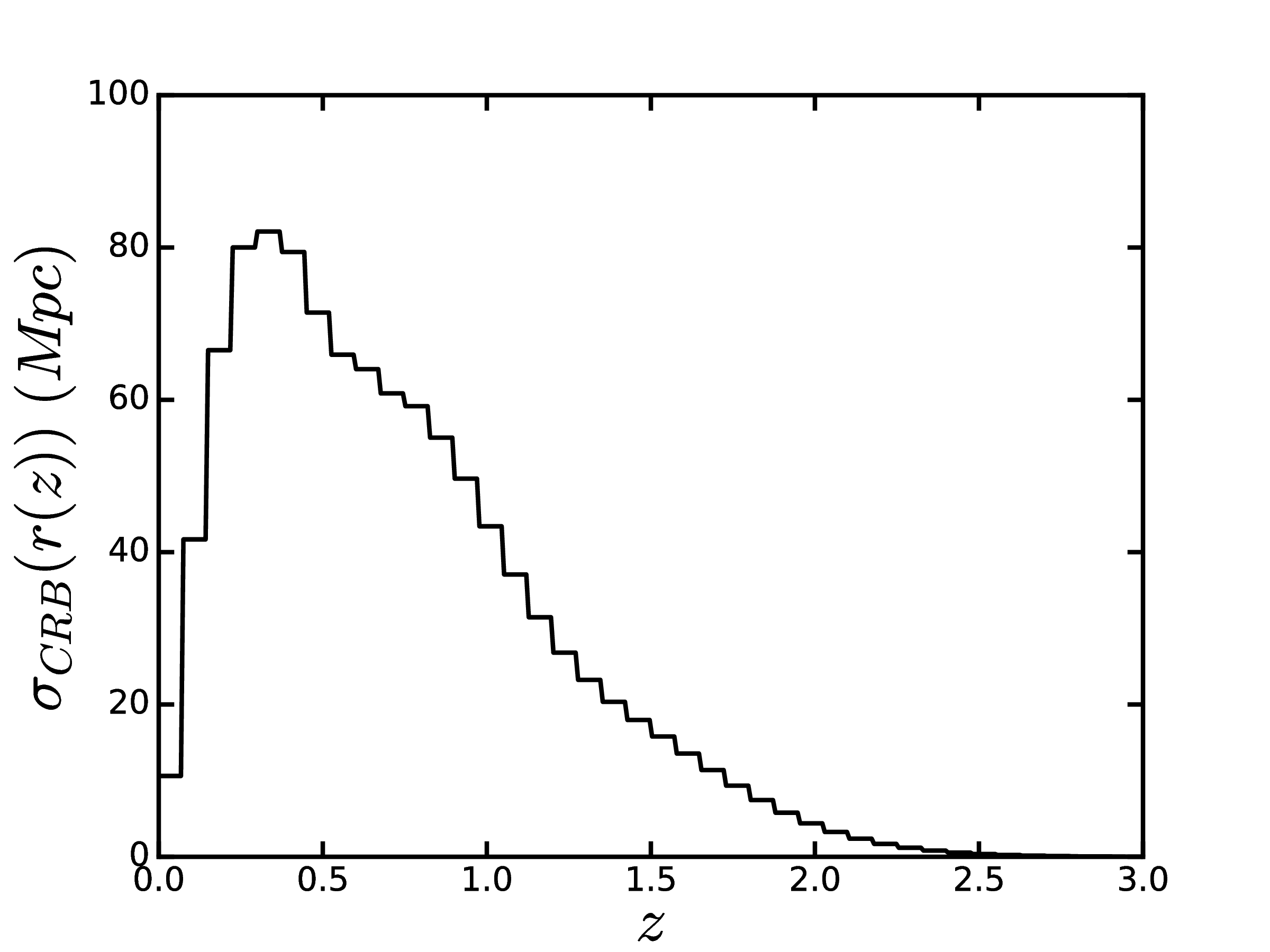}
        \includegraphics[width=5.5cm]{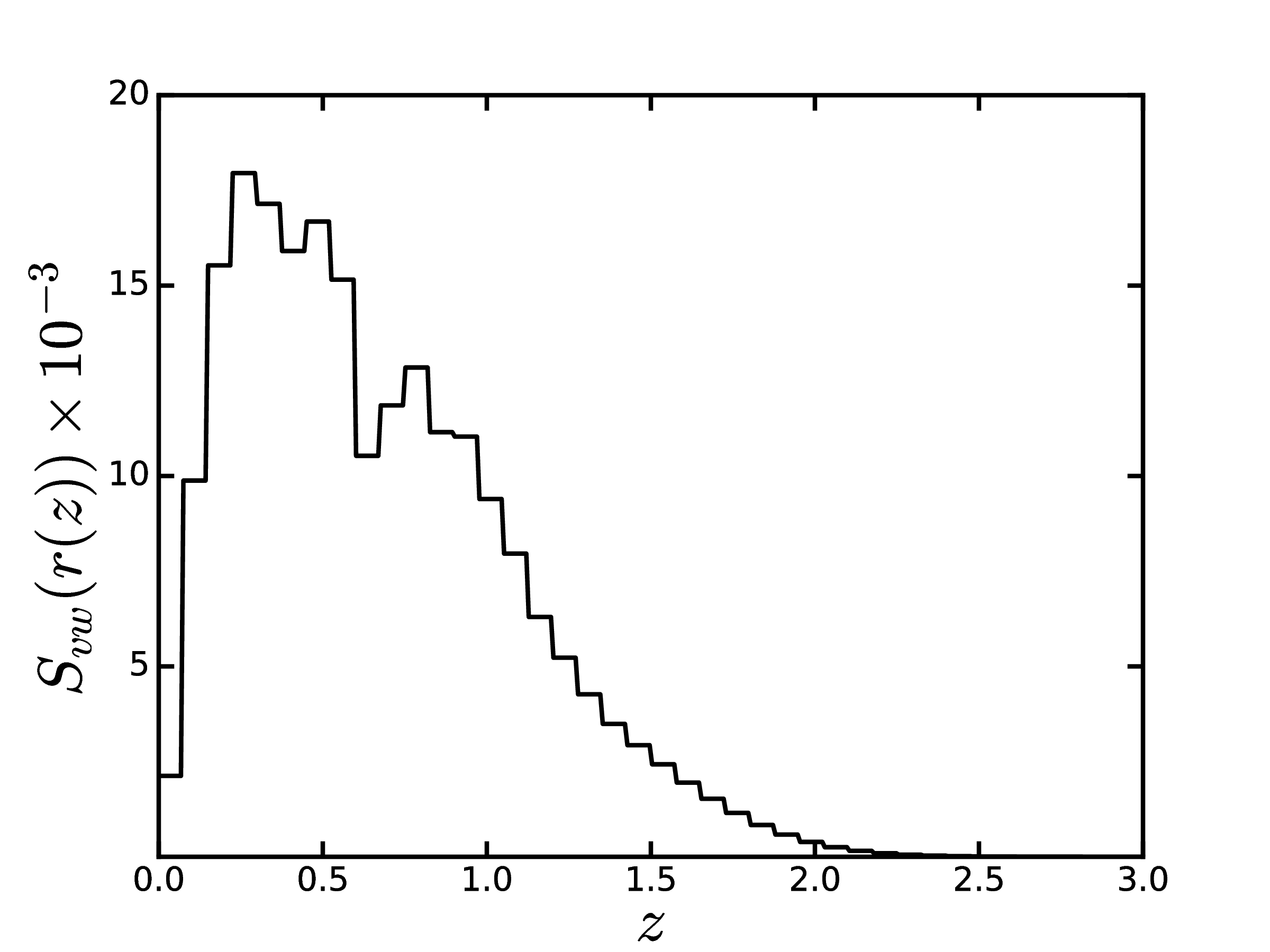}
        \includegraphics[width=5.5cm]{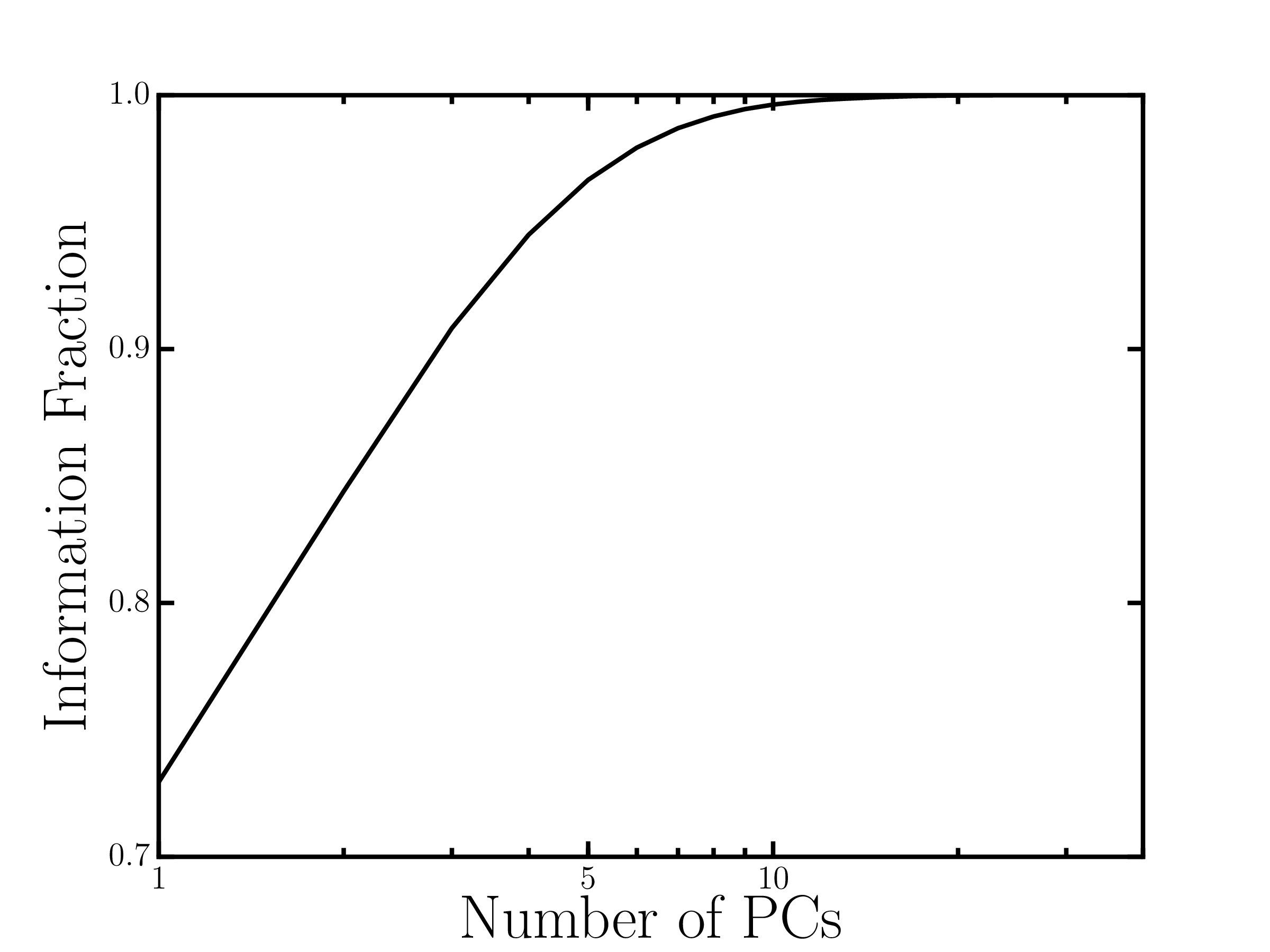}
		\caption{The first $2$ super tomographic lensing kernel principal components, the transverse co-moving distance for the fiducial cosmology, the Cramer-Rao bound, the variance weighted sum and the information fraction for the number of principal components. See Section~\ref{sec:pca_guide} for an explanation of the terminology. From $S_{vw}$ and the Cramer-Rao Bound, this method is primarily sensitive to $z \in \left[0., 1.5 \right].$ Three PCs capture the majority of the information.}
		\label{fig:tomo_kernel_fig}
	\end{minipage}
\end{figure*}

\begin{table*}[t]
  \centering
    \caption{Relative information content, $I_{tot}$, (see equation \ref{eq:info}) for the lensing kernel and the power spectrum for $10$-bin tomography with different $\ell$-cuts. The maximum values of $\ell$ is denoted as $\ell_{max}$. For cuts below $\ell = 1000$, $\sim 50 \%$ of the information is lost}
  \begin{tabular}{l|c@{\hskip 18pt}c@{\hskip 18pt}c@{\hskip 18pt}c@{\hskip 18pt}c@{\hskip 18pt}r}
  \hline \hline
   {\bf $\ell_\text{max}$} & $3000$ & $2500$ & $2000$ & $1500$ & $1000$ & $500$\\
    \hline
    {\bf $I_{tot}$ (lensing kernel)}  & $1.00$ &$0.94$ & $0.86$ & $0.75$ & $0.60$ & $0.38$\\
    {\bf $I_{tot}$ (power spectrum)} &  $1.00$ & $0.94$ & $0.84$ & $0.71$ & $0.51$ & $0.26$ \\
    \hline \hline
  \end{tabular}
  \label{tab:2}
\end{table*}

   \begin{figure}
   \centering
    \vspace{2mm}
    \includegraphics[width=85mm]{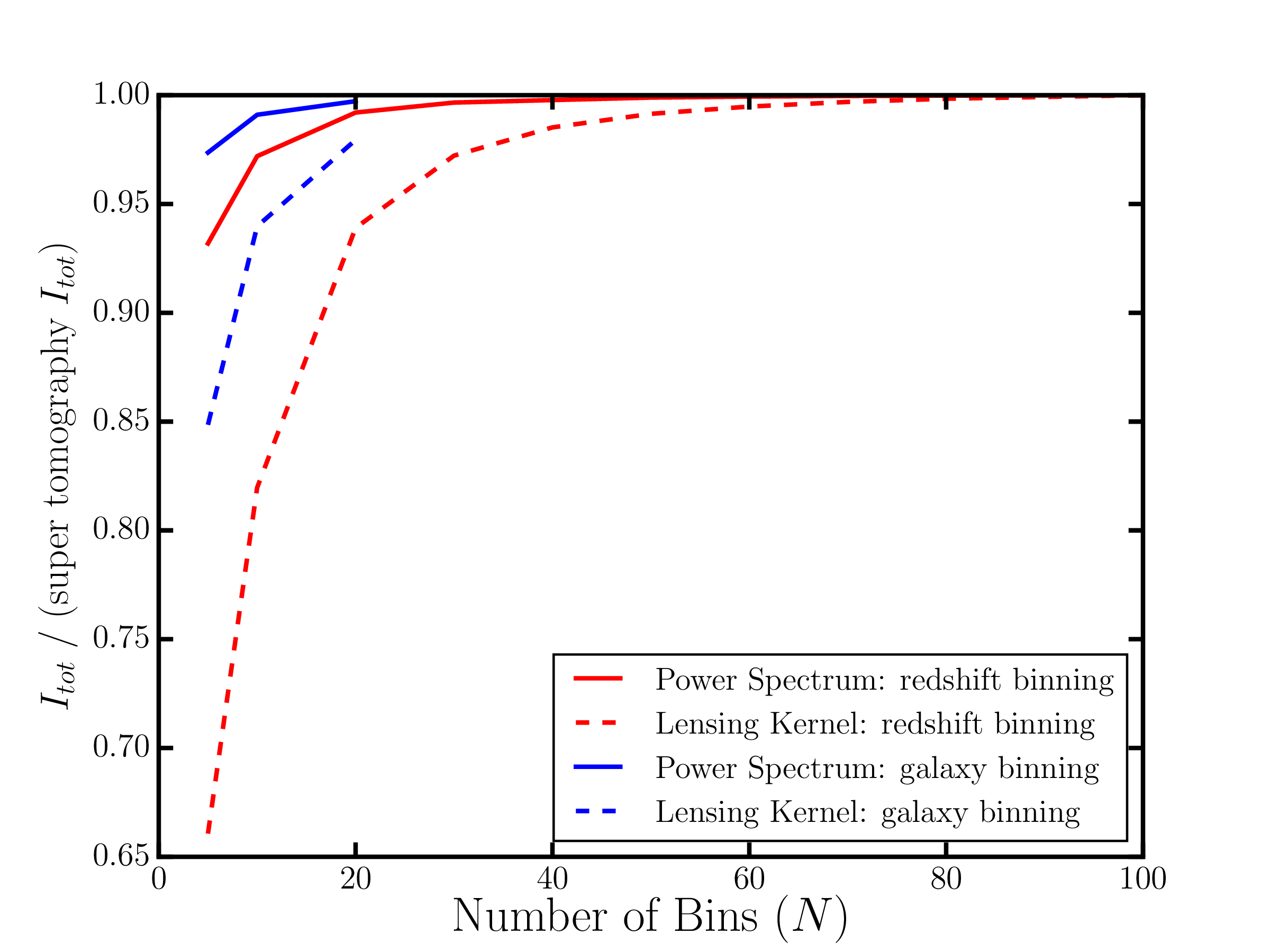}
    \caption{Fraction of the total information content, $I_{tot}$, of super tomography for different binning strategies as a function of the number of tomographic bins. {\bf Red:} Equally spaced bins in redshift. {\bf Blue:} Equal number of galaxies per bin. {\bf Solid Lines:} Power spectrum information fraction. {\bf Dashed Lines:} Lensing kernel information fraction. Equal number of galaxies per bin initially converges faster, but this binning scheme loses information at high-$z$ and is intractable for a large number of bins (see Section~\ref{sec:binning}).  $99\% (99.9 \%)$ of the information is captured from both the lensing kernel and the power spectrum with $50(90)$ bins.}
   \label{fig:bin_convergence}
    \end{figure}

   \begin{figure}
   \centering
    \vspace{2mm}
    \includegraphics[width=85mm]{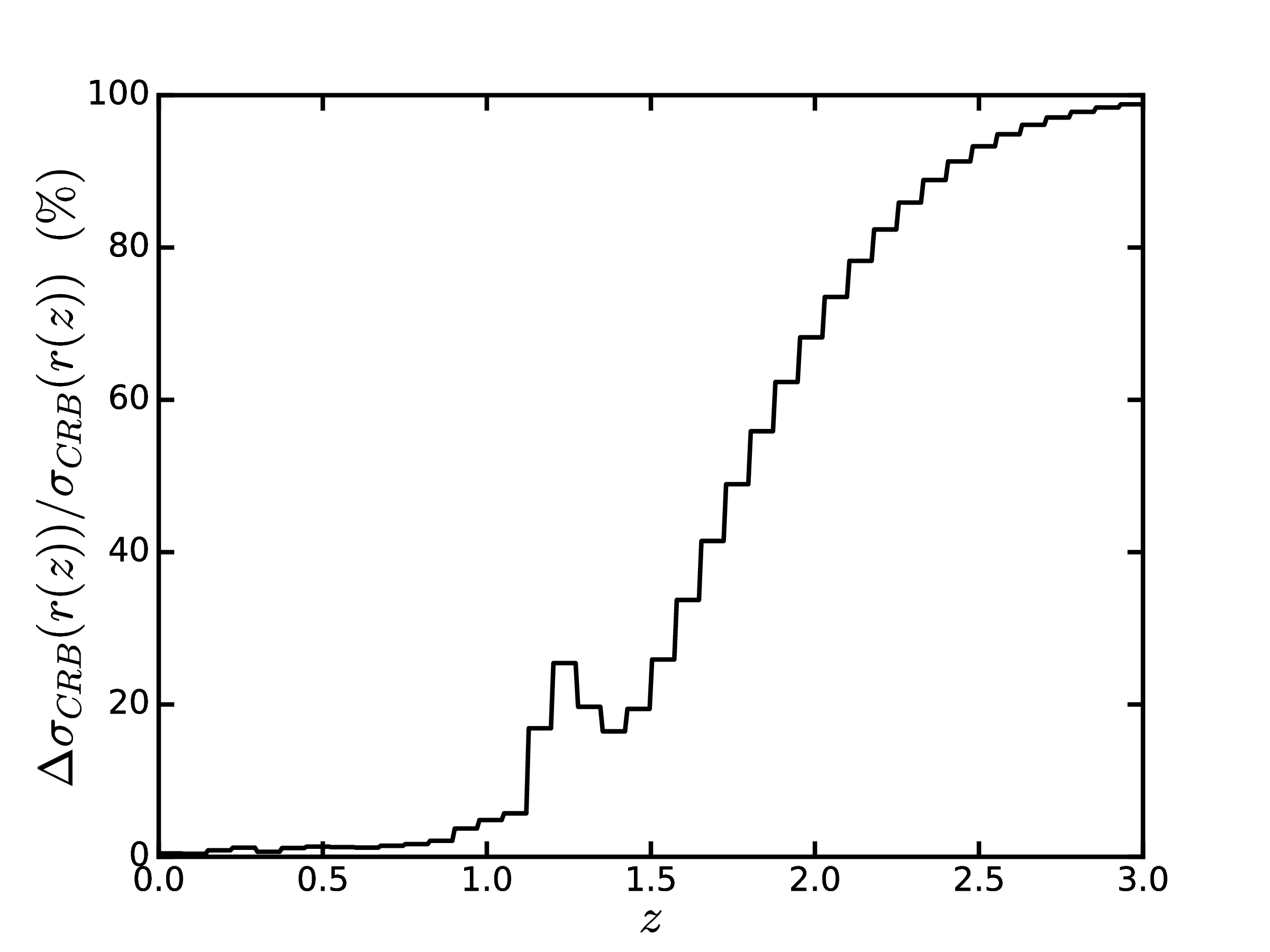}
    \caption{The relative difference in the Cramer-Rao bounds, on the lensing kernel, between super tomography and $10$-bin tomography with an equal number of galaxies per bin. $10$-bin tomography primarily loses information at high-$z$ (see Section~\ref{sec:binning}). At redshifts above $z= 2$, more than half the lensing kernel information is lost in the $10$-bin case. The difference in the total information content extracted remains small since most of the information comes from redshifts below $z = 1.5$. The small peak in the plot at $z= 1.3$ is due to coarseness of $10$-bin tomography and we have checked that this feature disappears when more bins are used.}
   \label{fig:tomo_kernel_more_bins_fig}
    \end{figure}

   \begin{figure}
   \centering
    \vspace{2mm}
    \includegraphics[width=85mm]{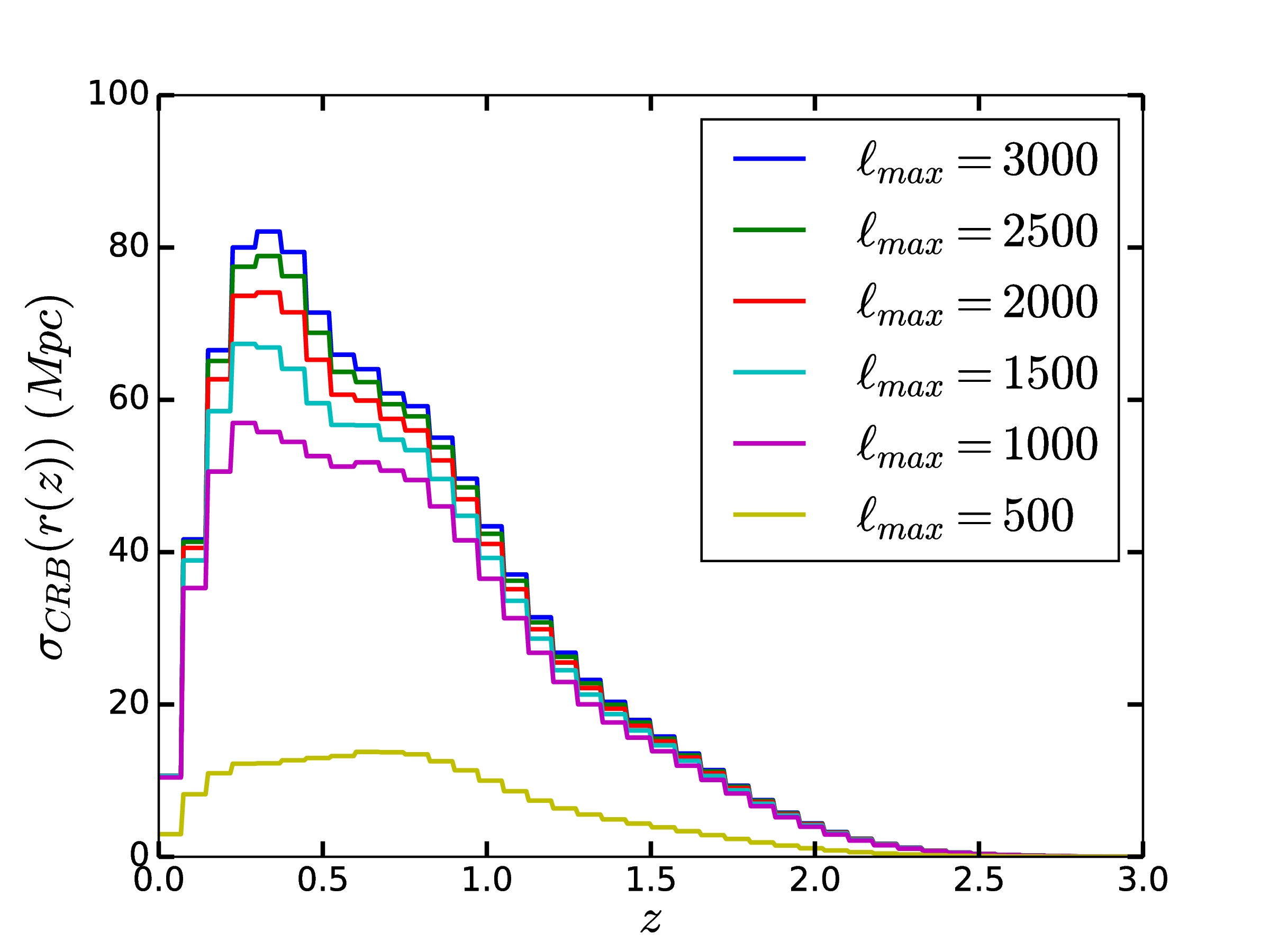}
    \caption{The super tomographic Cramer-Rao bound on the lensing kernel with different $\ell$-cuts. Information is primarily lost intermediate redshifts, near $z=0.5$, and catastrophically lost below $\ell_{max} < 1000.$}
    \label{fig:kernel_l_cut_compare}
    \end{figure}

We determine the lensing kernel PCs for tomography with an equal number of galaxies per bin, tomography with equally spaced $z$-bins and 3D cosmic shear with $\ell_{\rm max} = 3000$. We will refer to equally spaced $z$-bin tomography with $100$ bins as \textit{super tomography}. The PCAs for super tomography with smaller $\ell$-mode cuts are also found. Our modelling choices (see Appendix~\ref{sec:model choice}) mimic the Euclid wide survey. The PCs are found in $40$ redshift slices with $z \in \left( 0., 3.0 \right].$

\par The super tomographic PCs are shown in Figure~\ref{fig:tomo_kernel_fig}, along with the the fiducial co-moving distance which is actually being constrained, the Cramer-Rao Bound, the Variance Weighted Sum and a plot of the fraction of the information content captured by the first $N$ PCs. From the Cramer-Rao Bound and variance weighted sum it is clear that Euclid will primarily be sensitive to the lensing kernel for redshifts in the range $z \in (0.1,1.0)$. Also, only $3$ PCs are needed to capture $92 \%$ of the  the information.
\par The PCs for $10$-bin tomography with an equal number of galaxies per bin and 3D cosmic shear look nearly identical to the super tomographic ones plotted in Figure~\ref{fig:tomo_kernel_fig}. However slightly less information is captured for both the other analyses. This is summarised in Table~\ref{tab:1} {\bf Case I}.
\par 3D cosmic shear should capture as much information as super tomography since no radial data compression takes place. This is not the case in our analysis, where 3D cosmic shear captures $6 \%$ less information than super tomography. This is because in order to calculate the lensing spectra quickly, we use a low resolution computation grid. In the Appendix~\ref{sec:convergence checks}, we investigate using a higher resolution computation grid using lower resolution PCs. At the highest resolution we considered, only $3 \%$ of the information is lost to numerical noise (Table~\ref{tab:1} {\bf Case II}).
\par In $10$-bin tomography $6 \%$ of the lensing kernel information is lost due to inherent data compression. Using more bins would help reduce this number and we examine the convergence of the total information content for different binning strategies in Figure~\ref{fig:bin_convergence}. Initially an equal number of galaxies per bin converges more quickly, but it is infeasible for a large number of bins and information is lost at high redshifts (see Section~\ref{sec:binning}). Meanwhile an equal redshift spacing binning strategy captures $99\% (99.9 \%)$ with $50(90)$ bins. 
\par The relative difference in the Cramer-Rao Bound between tomography and super tomography is shown in Figure~\ref{fig:tomo_kernel_more_bins_fig}. This confirms that $10$-bin tomography loses information at high redshifts, beyond $z=1$, as expected. This difference is unimportant for constraining dark energy equation of state parameters -- that only become relevant below $z \sim 1$, explaining the quick convergence of dark energy constraints with the number of tomographic bins found in \cite{bridle2007dark}. However, these higher redshifts are precisely where we expect a cross-correlation signal with CMB lensing. Cross-correlating with the CMB lensing signal will help bring lensing systematics under control \cite{merkel2017parameter} to substantially improve the dark energy and neutrino mass Figure of Merit \cite{kitching20143d}.
\par Finally we assess the impact of angular scale cuts. Taking $\ell$-cuts reduces the sensitivity to the lensing kernel. This is shown in Table~\ref{tab:2}. When taking $\ell$-cuts, information is primarily lost at intermediate redshifts, near $z=0.5$. This can be seen from Figure~\ref{fig:kernel_l_cut_compare}, where the Cramer-Rao Bounds on the lensing kernel for different $\ell$-cuts are plotted.

 \subsection{Power Spectrum PCs} \label{power spec}

\begin{figure*}
	\begin{minipage}{1.0\textwidth}
		\includegraphics[width=5.5cm]{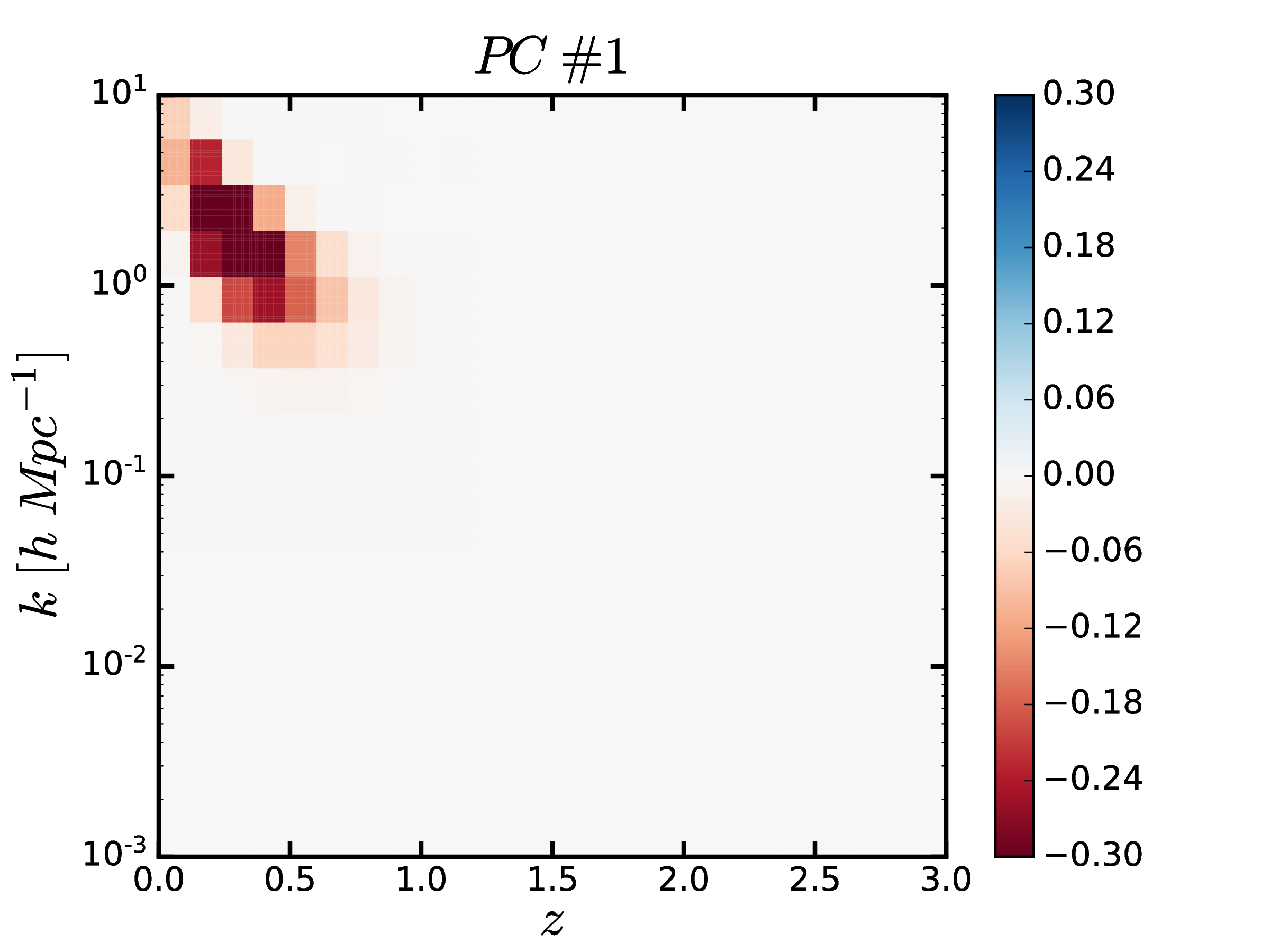}
		\includegraphics[width=5.5cm]{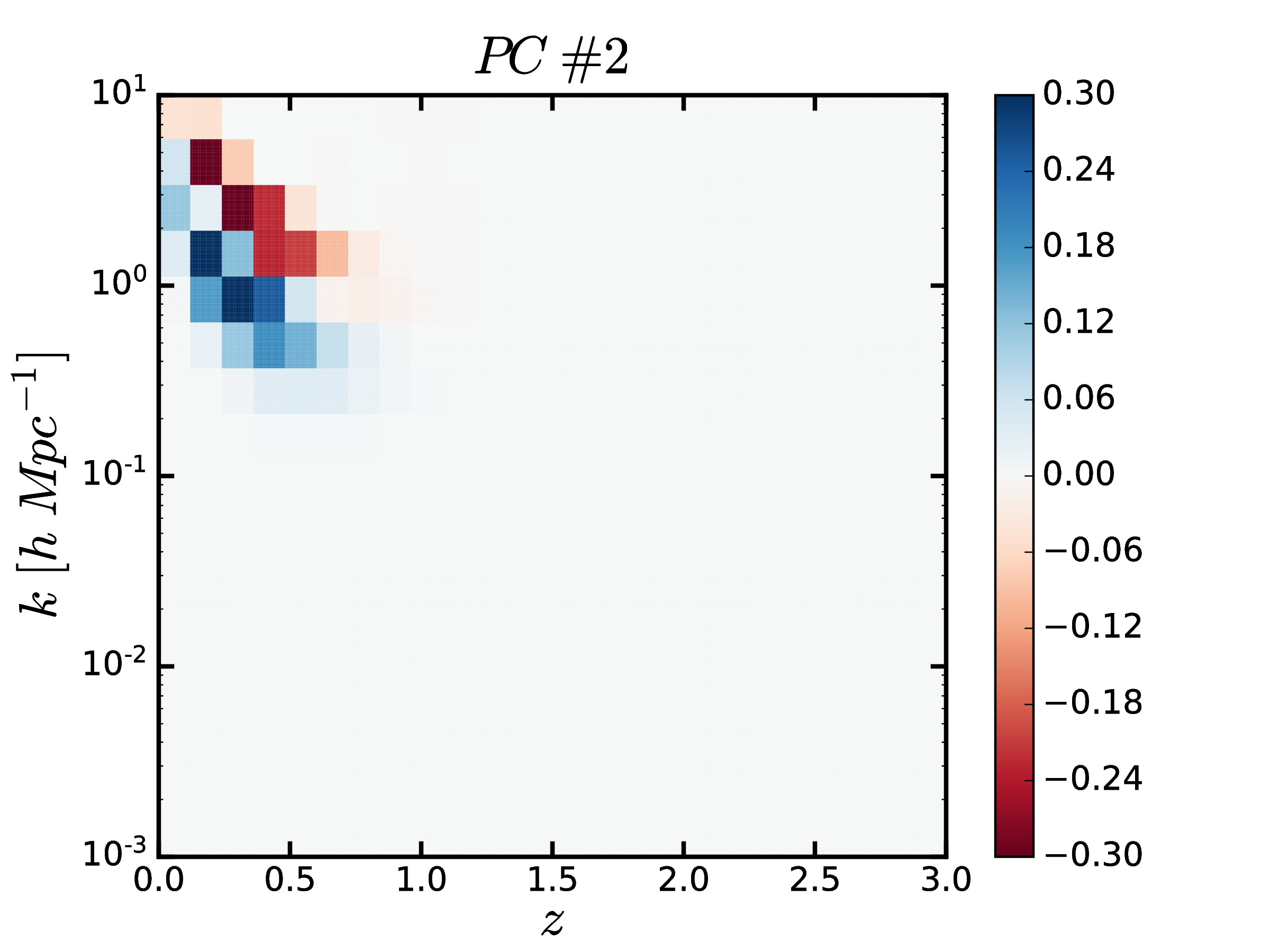}
		\includegraphics[width=5.5cm]{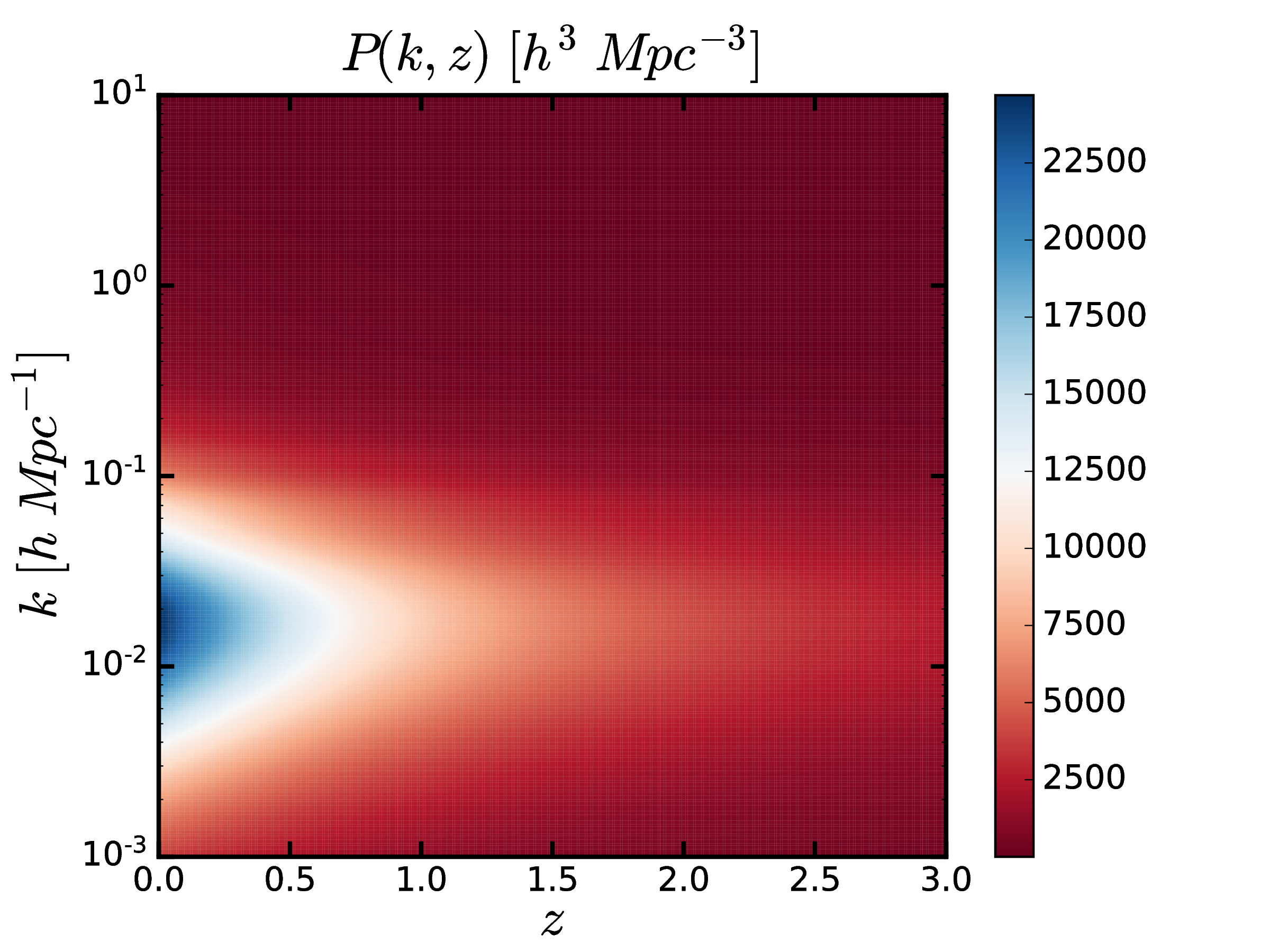}
        \includegraphics[width=5.5cm]{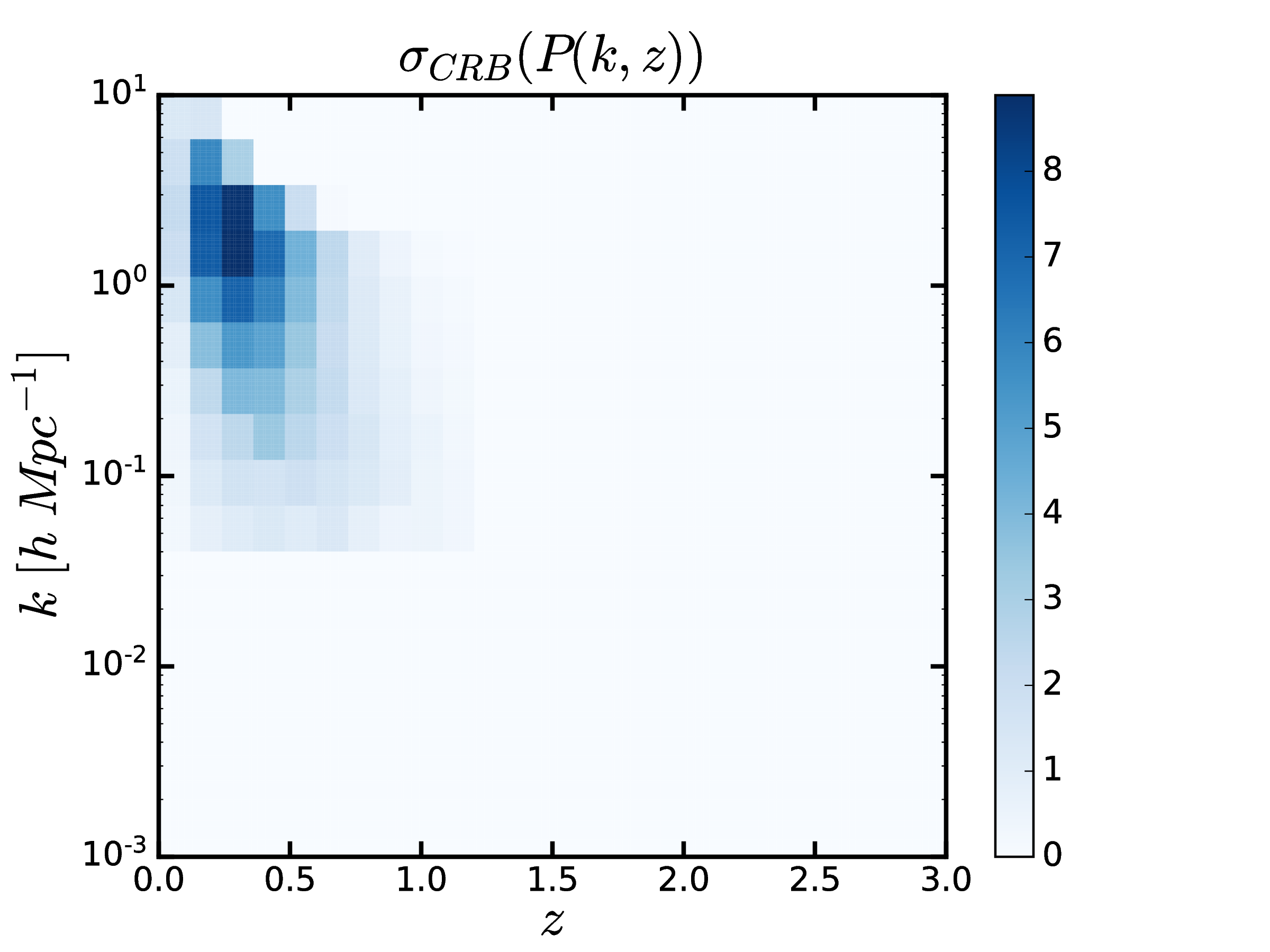}
        \includegraphics[width=5.5cm]{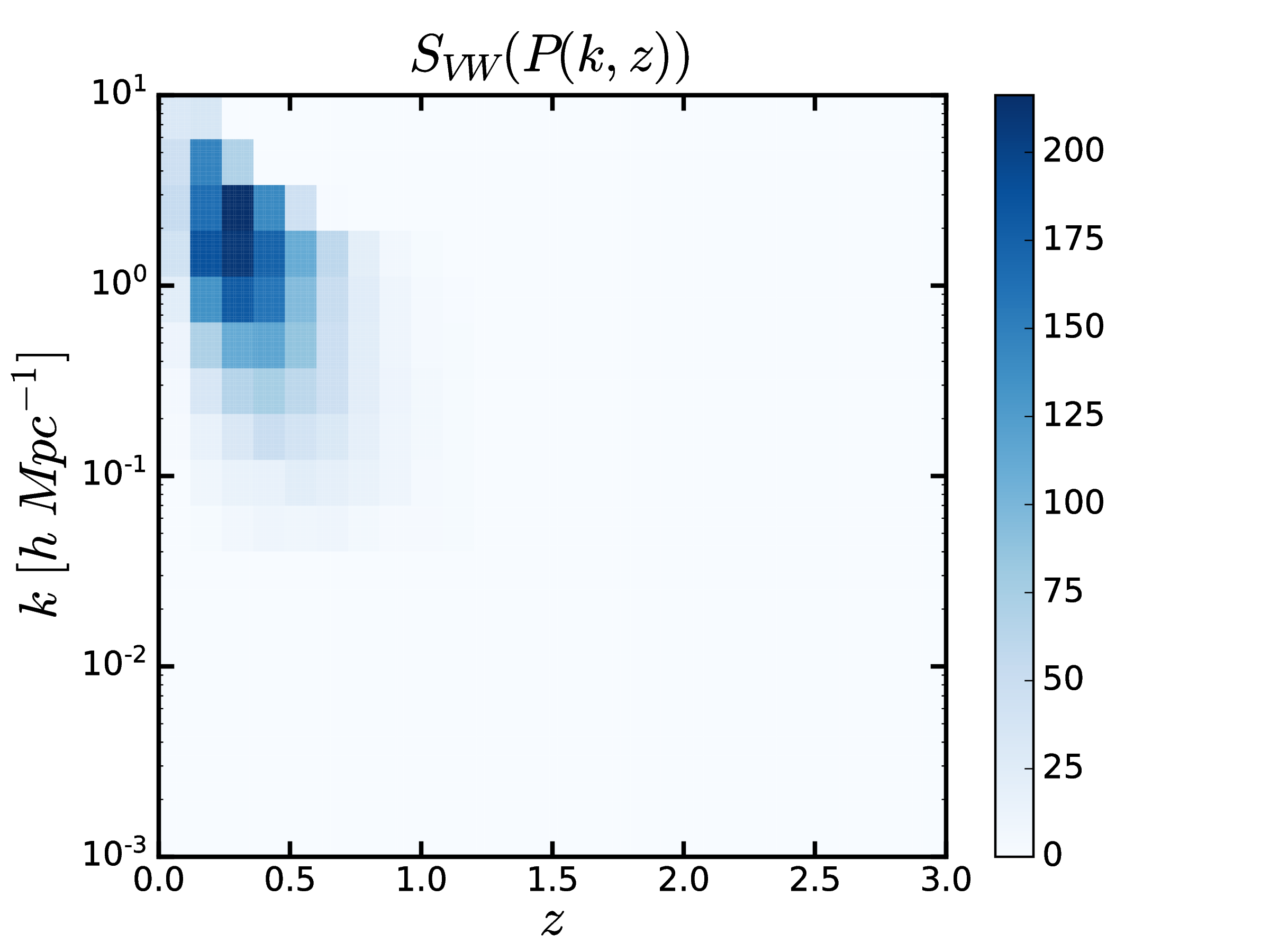}
        \includegraphics[width=5.5cm]{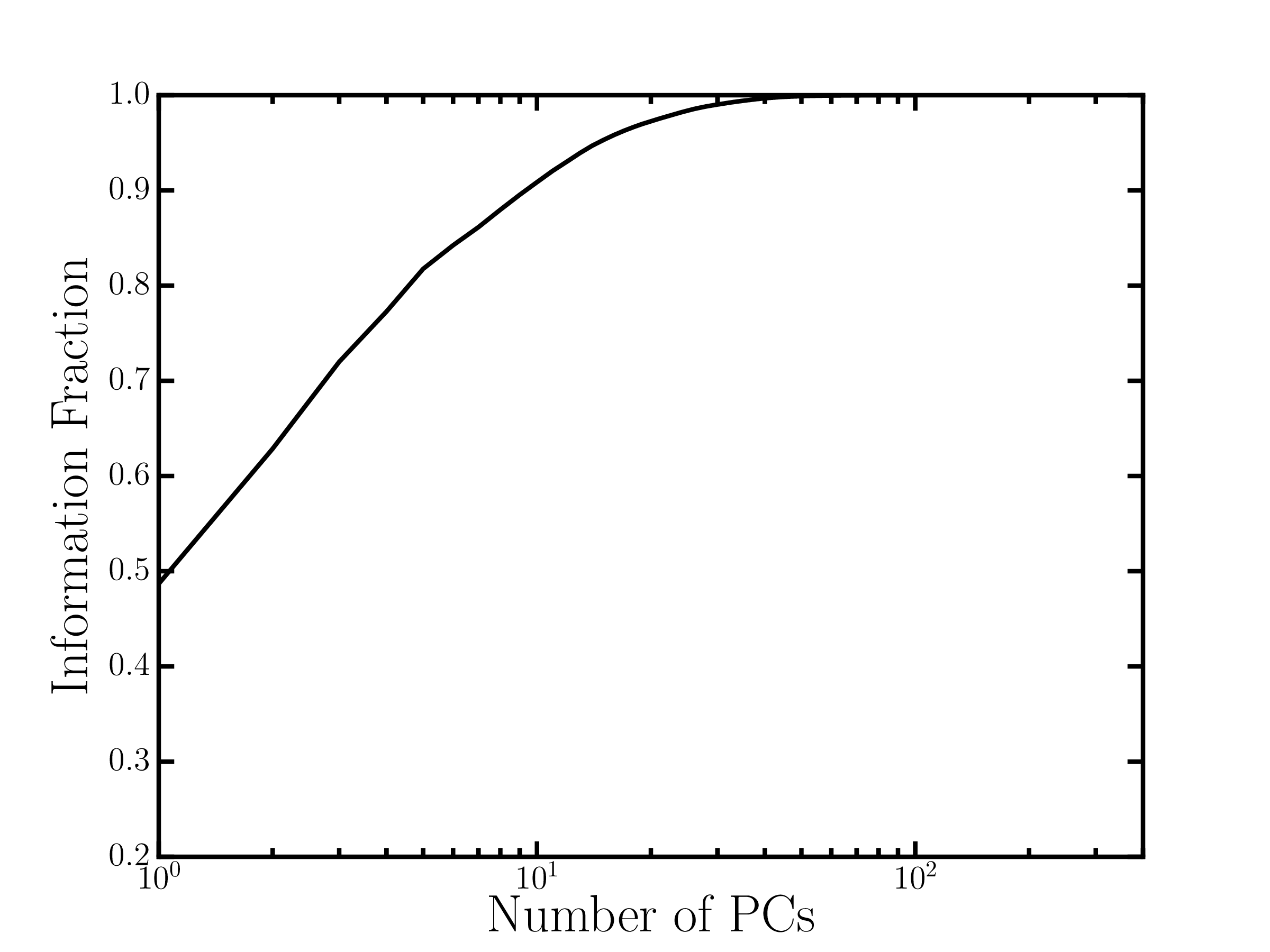}
		\caption{The first two super tomographic lensing power spectrum PCs, the power spectrum of the fiducial cosmology, the Cramer-Rao bound, the variance weighted sum and the information fraction. See Section~\ref{sec:pca_guide} for an explanation of the terminology. Cosmic shear is primarily sensitive to the power spectrum at large scales above 0.5 $h \text{Mpc} ^{-1}$ and at low redshift. Nearly half the signal lies above $k = 1.5 h \text{ Mpc} ^{-1}$ where the power spectrum becomes hard to model. Within principal components, bins are strongly correlated across this cut so power spectrum modelling errors at high-$k$ induces bias at low $k$. }
		\label{fig:tomo_fig1}
	\end{minipage}
\end{figure*}

   \begin{figure}
   \centering

    \vspace{2mm}
    \includegraphics[width=85mm]{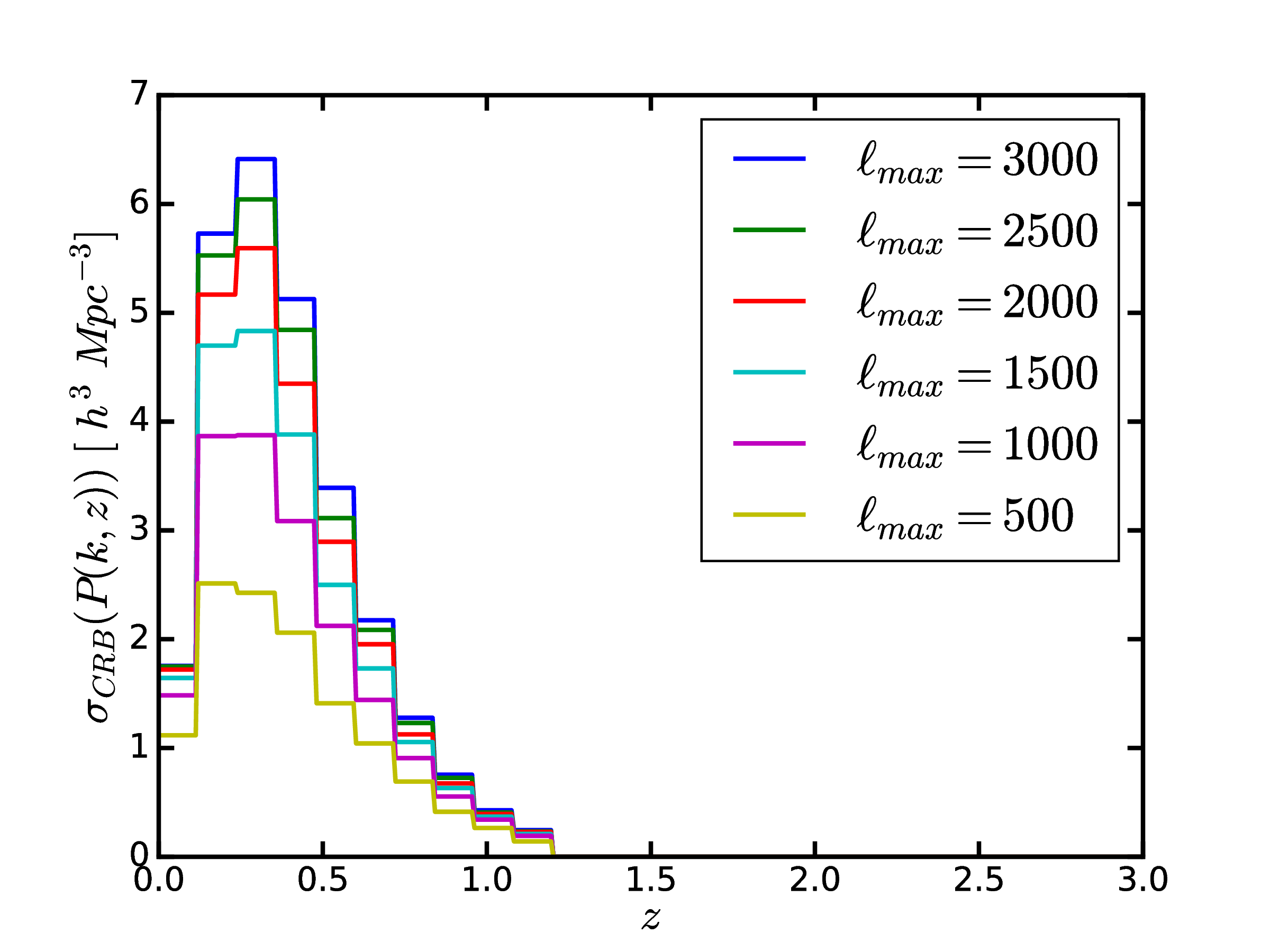}
    \includegraphics[width=85mm]{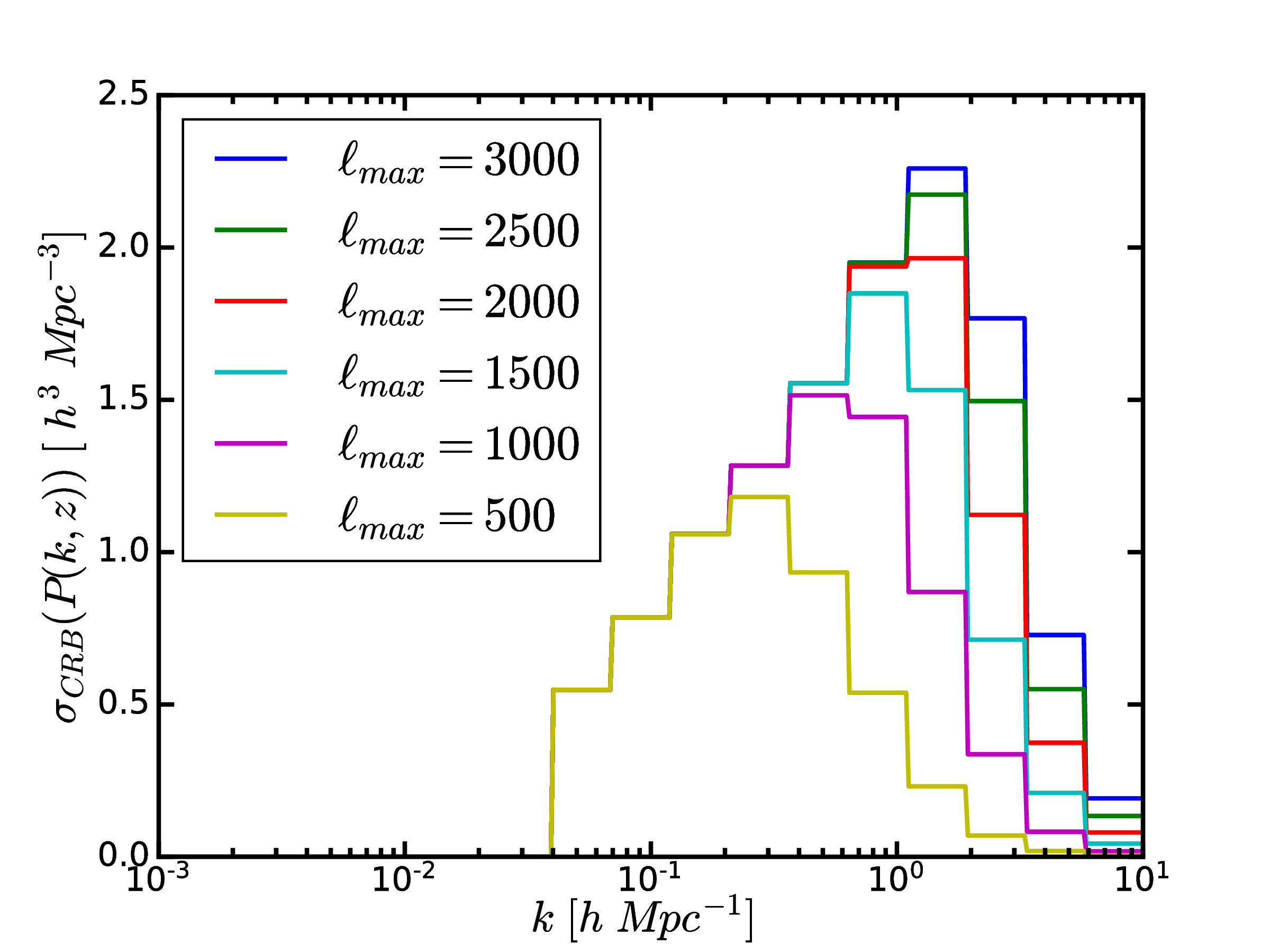}
    \caption{The average Cramer-Rao projected onto the $z$-axes (above) and $k$-axis (below) for different $\ell$-mode cuts in $10$-bin tomography. Cuts below $\ell \sim 1000$ significantly reduce the sensitivity to $k$-modes above $1.5 h \text{ Mpc} ^ {-1}$. }
    \label{fig:pca_l_cuts}
    \end{figure}

\begin{figure*}
	\begin{minipage}{1.0\textwidth}
		\includegraphics[width=5.5cm]{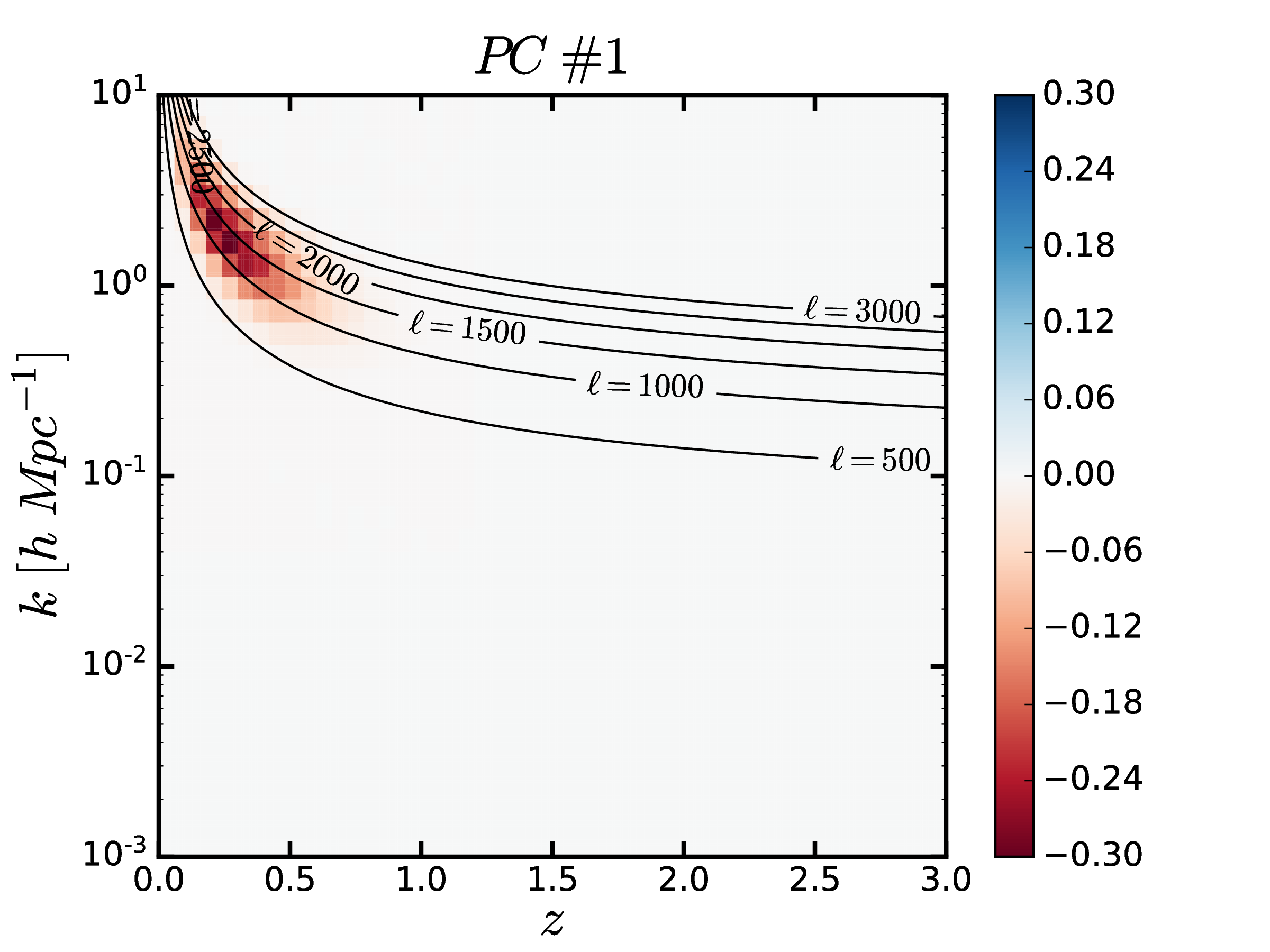}
		\includegraphics[width=5.5cm]{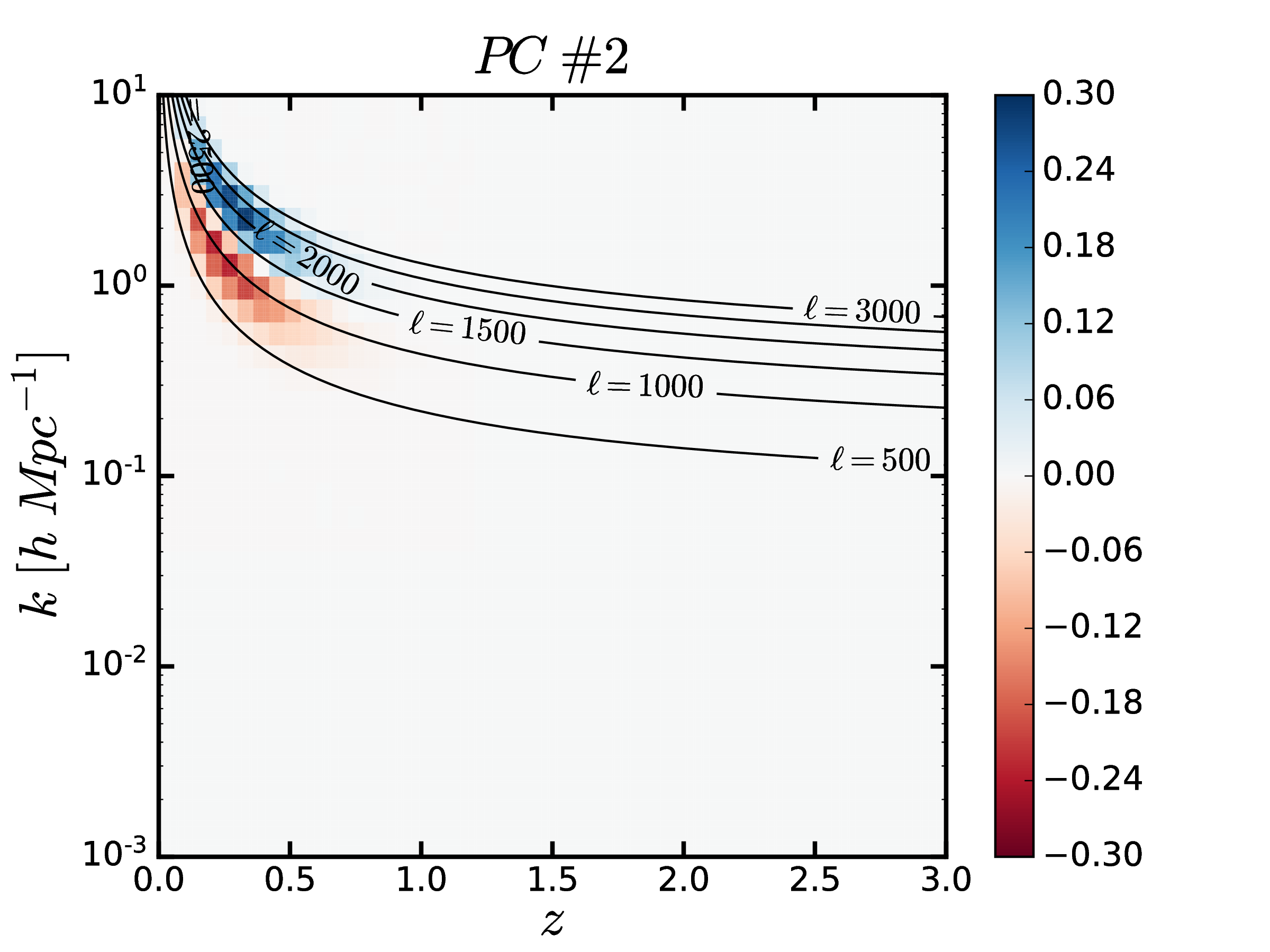}
		\includegraphics[width=5.5cm]{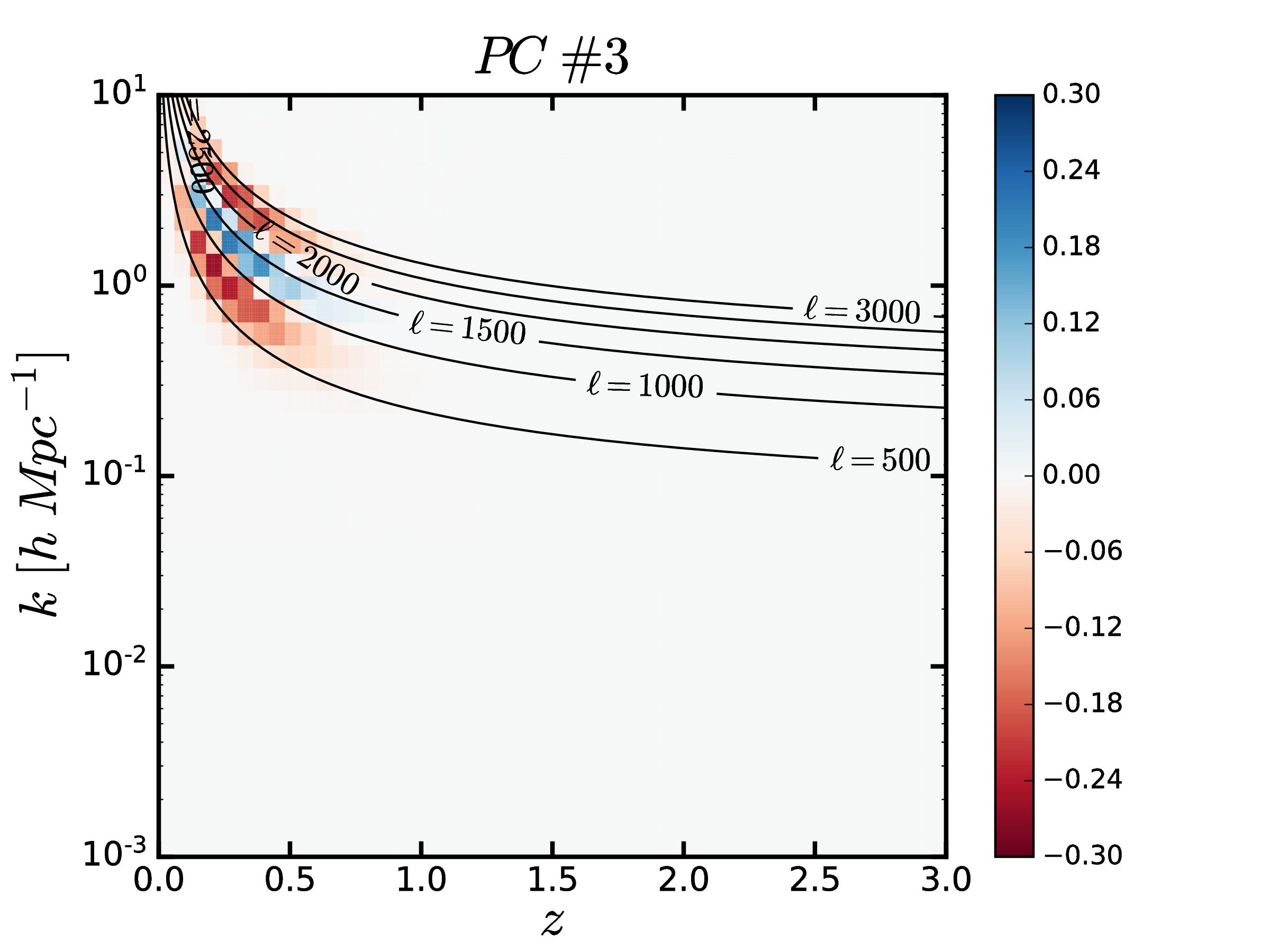}
        \includegraphics[width=5.5cm]{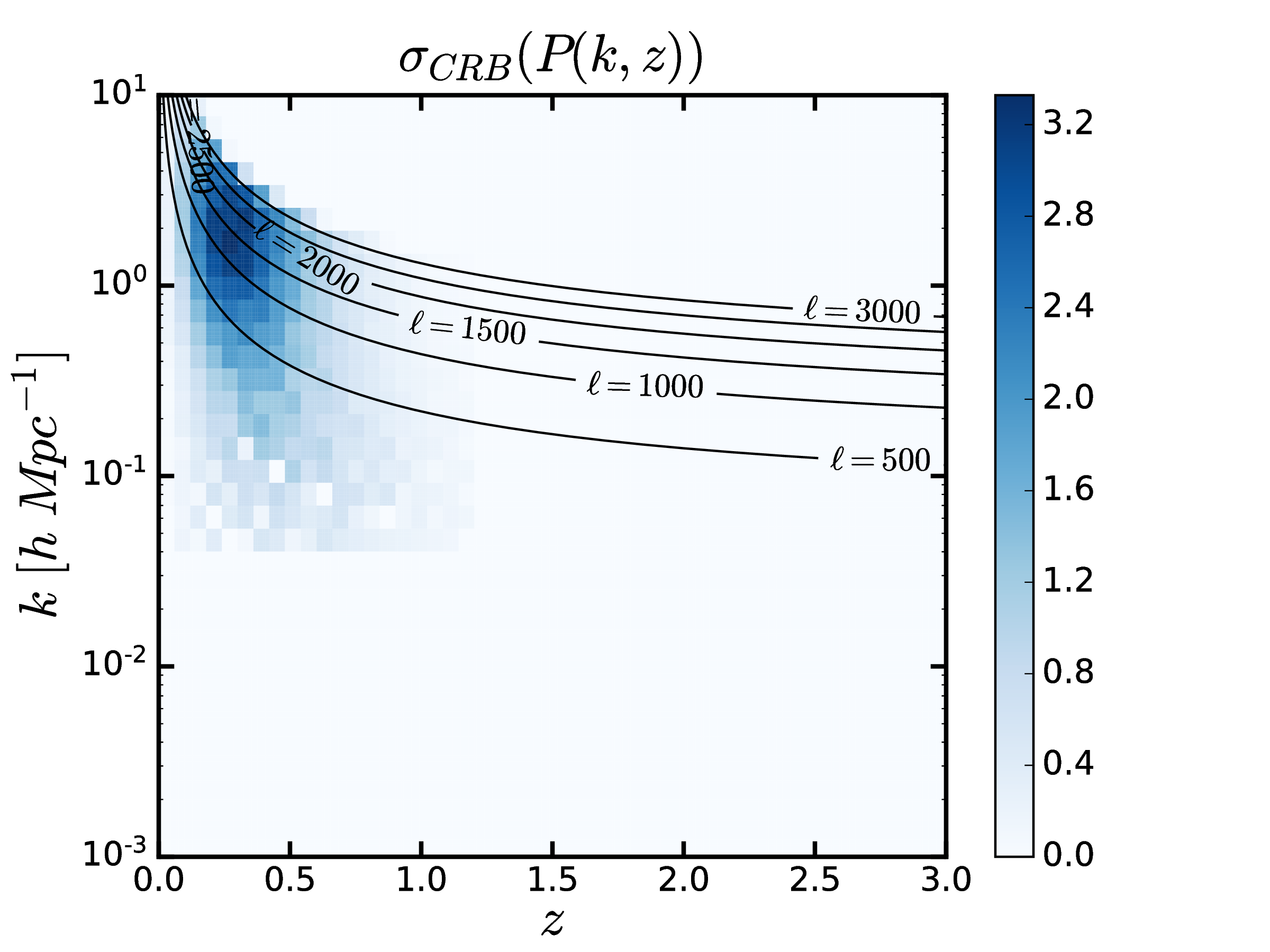}
        \includegraphics[width=5.5cm]{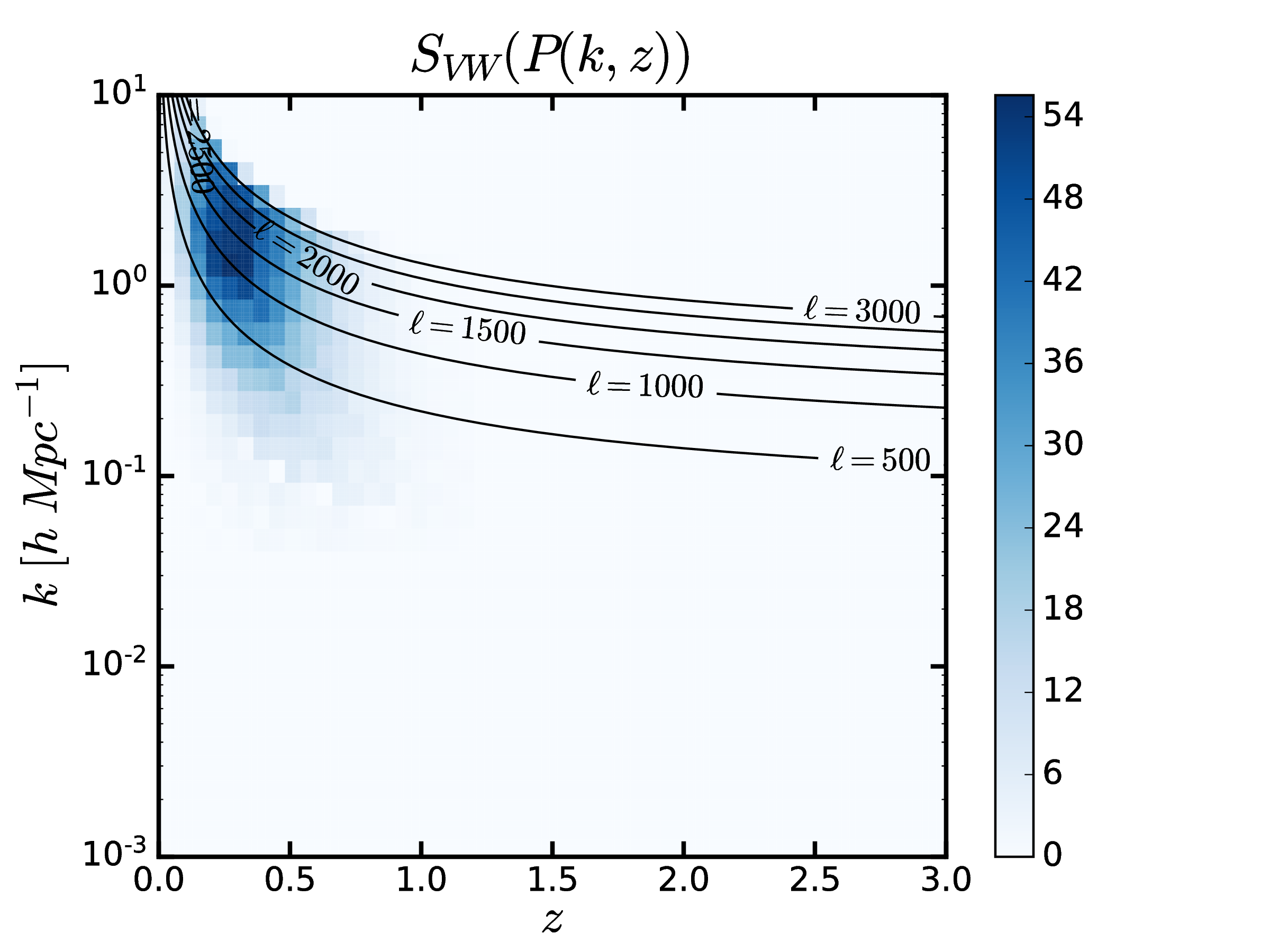}
        \includegraphics[width=5.5cm]{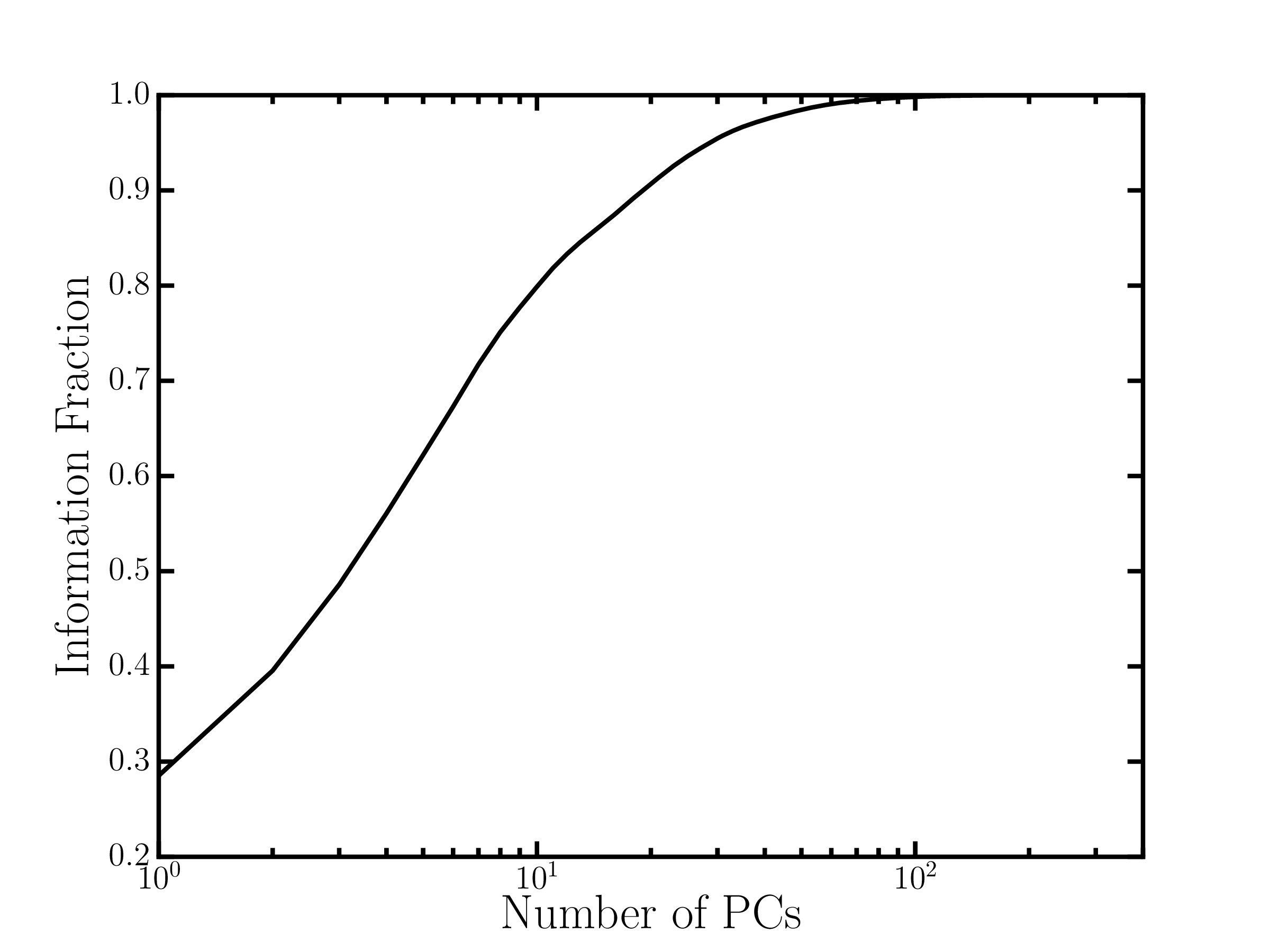}
		\caption{The first three high-resolution super tomographic power spectrum PCs, the Cramer-Rao bound, the variance weighted sum and the information fraction. The PCs follow Limber lines (see discussion below equation \ref{eq:limber}) plotted for different $\ell$-modes in black.  }
		\label{fig:tomo_fig_high_res}
	\end{minipage}
\end{figure*}

We compute the power spectrum PCs for tomography with an equal number of galaxies per bin, tomography with equally spaced $z$-bins and 3D cosmic shear with $\ell_{\rm max} = 3000$. PCs for super tomography with lower $\ell$-cuts are also found. To save time, we compute these on a coarse grid before zooming in on the region of primary sensitivity by first perturbing the matter power spectrum on a $5 \times 5$ grid logarithmically spaced in $k$ and linearly spaced in $z$. More than $95 \%$ of the signal is contained in the first two $z$ bins and the last two $k$ bins. The PCs are then computed on a $10 \times 10 $ grid just inside this region. 
\par The resulting PCs for super tomography are shown in Figure \ref{fig:tomo_fig1} and the total information content is displayed in Table \ref{tab:1} (Case I). A high resolution super tomography PCA run, on a $20 \times 20$ grid, is plotted in Figure \ref{fig:tomo_fig_high_res}. Due to memory constraints a high-resolution run is not done for 3D cosmic shear. 
\par The super tomographic PCs look very similar to those of 3D cosmic shear (not shown), but since the eigenvalues are larger, more information is extracted (see Table \ref{tab:1}) in the former case, but just like in the lensing kernel case, this is due to the slow numerical convergence of 3D cosmic shear. The ratio of the information contents of 3D cosmic shear and super tomography is plotted in Figure~\ref{fig:convergence} on a two-by-two PCA grid. By a resolution of $N = 5000$, the relative information content of $3D$ cosmic shear is within $ 3 \%$ of super tomography. This is displayed in Table~\ref{tab:1} (Case II). 
\par The shape of the PCs is not surprising. The first principal component is a relatively broad feature following the Limber line $\ell \sim 1500$. Meanwhile the higher PCs show multiple broad features tracing multiple Limber lines. This is particularly noticeable for the higher resolution PCs in Figure~\ref{fig:tomo_fig_high_res}. In the absence of shot noise, each $\ell$-mode is independent and sensitive to the power spectrum along the Limber lines. However, shot noise induces correlation between Limber lines ($\ell$-modes), as the uncertainty on neighbouring bins in the power spectrum are correlated, causing the broad features in the PCs.  We have performed a test without shot noise, and the principal components trace separate Limber lines exactly, without broadening, as expected.
\par The lensing kernel ensures that cosmic shear experiments are most sensitive to regions of the matter power spectrum that are at half the comoving distance of the bulk of the source galaxies in the survey. Along with the temporally evolving amplitude of the power spectrum, this ensures that lensing is primarily sensitive to the power spectrum in the redshift range $z \in [0.1, 0.6]$ (see Figures~\ref{fig:tomo_fig1}).
\par The convergence of total information content with different binning strategies is plotted in Figure~\ref{fig:bin_convergence}. For an equal redshift spacing binning strategy, $99\% (99.9 \%)$ of the power spectrum information is captured in $20(60)$ bins.
\par Meanwhile around half of the signal to the power spectrum lies above $k = 1.5 h \text{ Mpc} ^{-1}.$ At such small scales the power spectrum is difficult to model, and there is usually a modelling error of around $10 \%$ \cite{mead2015accurate}. High and low-$k$ bins are correlated along the PCs, so a modelling error on the former will induce a bias in the later. We quantify the bias for different $\ell$-modes cuts in Section~\ref{sim section}.
\par Taking an $\ell$-mode cut removes sensitivity above a given $k$-cut. This is because low-$\ell$ Limber lines only lie above the $k$-cut at low redshift, where the sensitivity is suppressed by the lensing kernel, which peaks at half the distance to the peak of the galaxy distribution near $z = 0.7$. This is shown in Figure~\ref{fig:tomo_fig_high_res} where we project the inverse error onto the $z$ and $k$-axes, taking the average. Cuts below $\ell =1000$ significantly reduce the sensitivity to scales smaller than $k = 1.5 h \text{ Mpc} ^{-1},$ but around half the sensitivity to the power spectrum is lost (see also Table~\ref{tab:2}). Meanwhile the amount of information gained by included more $\ell$-modes slows rapidly above $\ell =2500$.

\subsection{Correlations in Power Spectrum Bias} \label{sec:sim req}

   \begin{figure}
   \centering
    \vspace{2mm}
    \includegraphics[width=85mm]{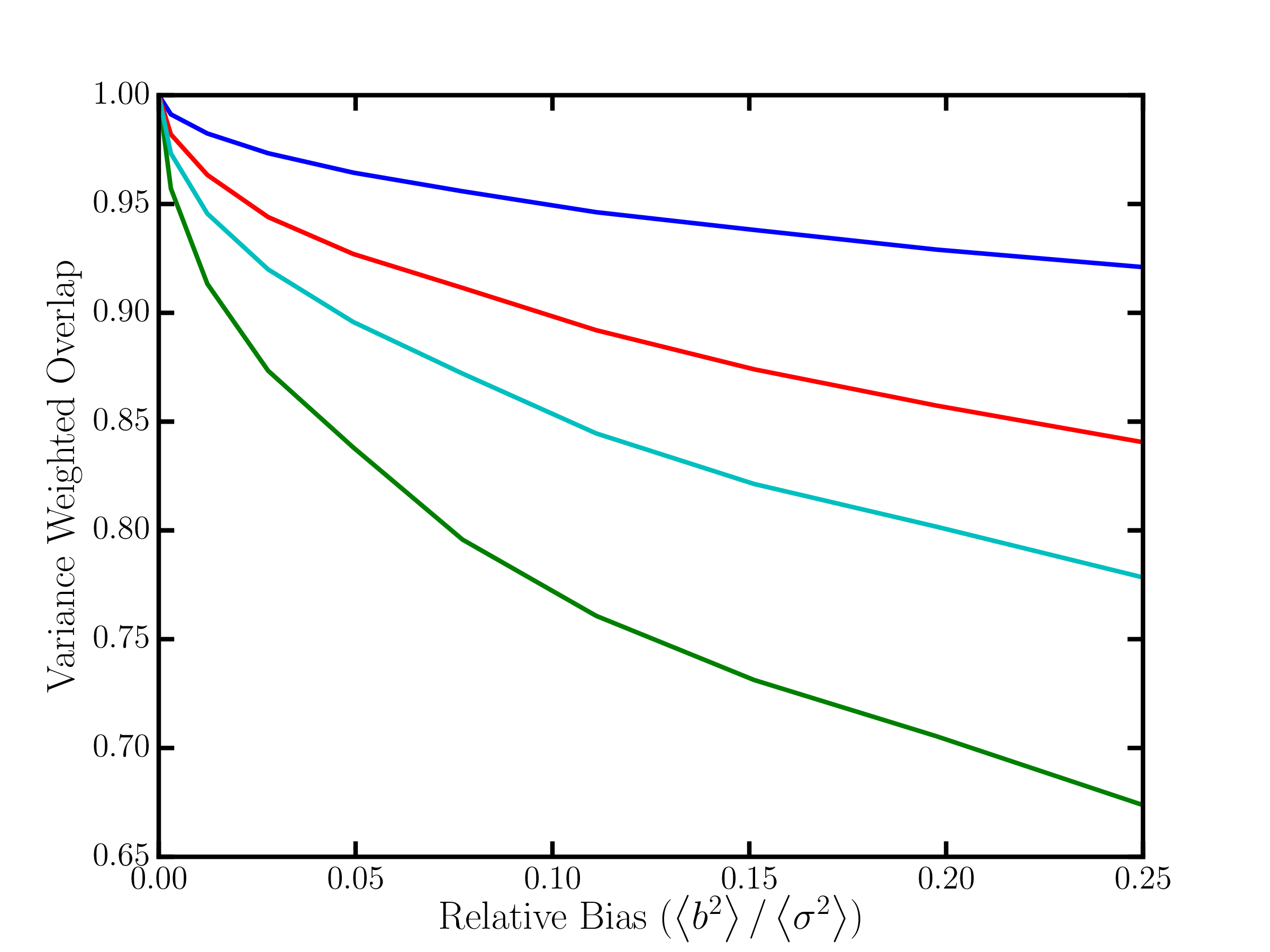}
    \caption{ $P_{VWO}$ for different covariance $K$. Correlation between bias in different $k$-$z$ regions has a large effect on simulation requirements. Factor of $3$ difference on $\langle b^2 \rangle / \langle \sigma^2 \rangle $ requirements for $P_{VWO}>0.95$.  \textbf{Extreme K:} \textbf{Blue}: $K$ is correlated with lensing PCs covariance $C$. \textbf{Green}: $K$ is anti-correlated with $C$. \textbf{Realistic K:} \textbf{Red}: $K$ is diagonal and constrains all PCs equally well. \textbf{Cyan}: Fiducial $K$-shape (see Figure~\ref{fig:requirements_plot_3}).}
    \label{fig:requirements_plot_1}
    \end{figure}

   \begin{figure}
   \centering
    \vspace{2mm}
    \includegraphics[width=85mm]{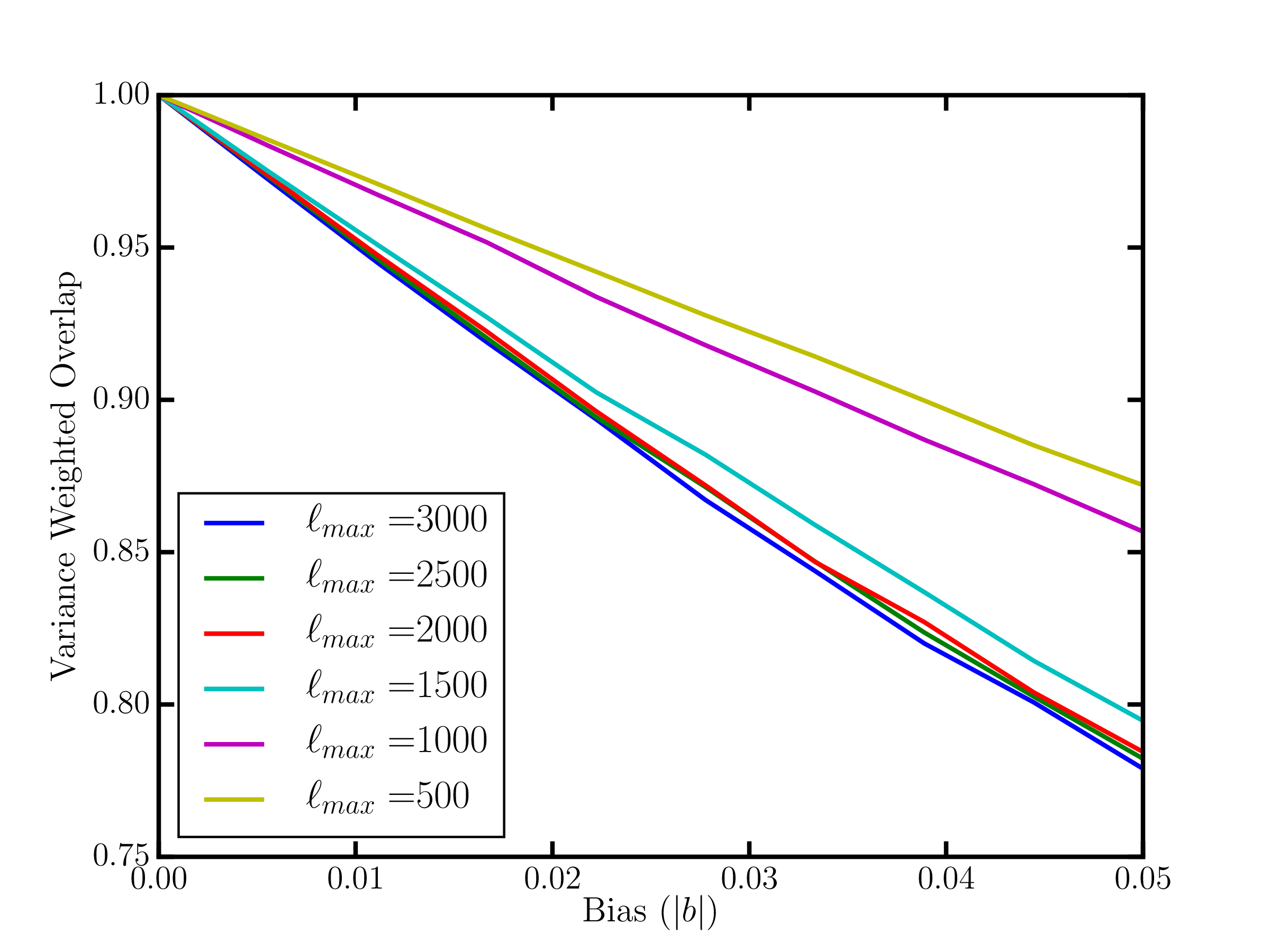}
    \caption{$P_{VWO}$ as a function of $|b|$ for different $\ell$-mode cuts. Bias is naively assumed to be maximally correlated in $k$ and constant everywhere. To ensure $P_{VWO} > 0.95$ for $\ell_{max} = 3000 (500)$ requires the bias to be less than $1 \% (3 \%)$. }
    \label{fig:requirements_plot_3}
    \end{figure}

   \begin{figure}
   \centering
    \includegraphics[width=85mm]{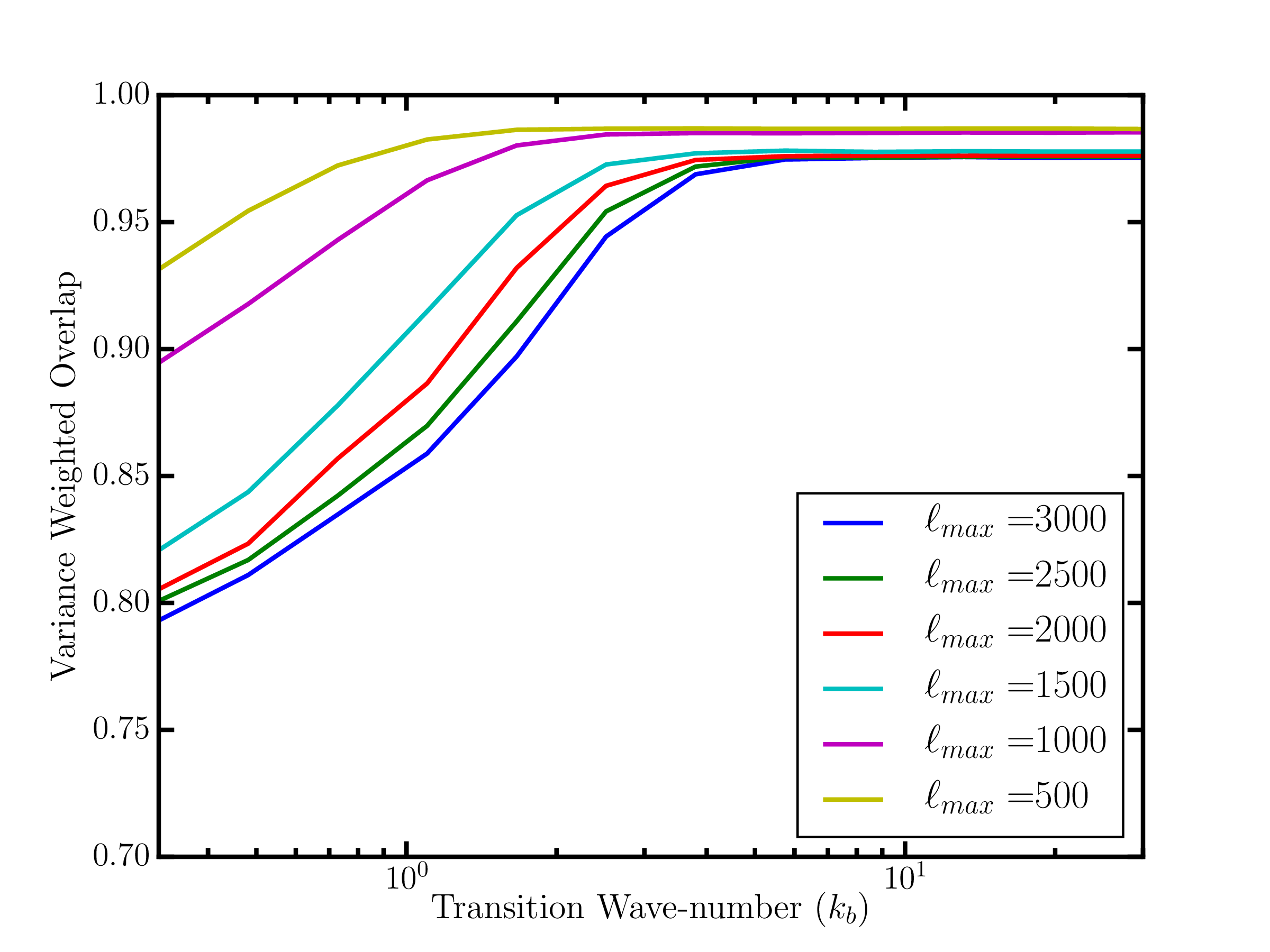}
    \caption{ $P_{VWO}$ assuming a $10 \%$ error above the transition wave-number $k = k_b$ and $1 \%$ below, for different $\ell$-cuts. In this scenario, to avoid taking an $\ell$-cut, the power spectrum should be known up to $k \sim 4 h \text{ Mpc} ^ {-1}$ for unbiased results. Increasing our knowledge of the power spectrum at scales smaller than $k \sim 7 h \text{ Mpc} ^ {-1}$ does not significantly change the bias on the lensing signal.}
    \label{fig:requirements_plot_4}
    \end{figure}

\par We use PCs for super tomography for the remainder of the Results section, since these fully capture the $3$D information.
\par It is clear that for cosmic shear studies not all regions of the matter power spectrum are equally important. Since lensing is barely sensitive to the power spectrum above $k = 10 h \text{ Mpc} ^ {-1}$, modelling errors at smaller scales can be large without inducing bias on parameters inferred from cosmic shear. Meanwhile accurately modelling the power spectrum near $k = 1.5 h \text{ Mpc} ^ {-1}$, where the sensitivity peaks, is extremely important. However there is a more subtle effect that can dramatically change requirement on the accuracy of power spectrum models. This is the degeneracy in modelling errors between regions in the power spectrum.  
\par  In Section~\ref{sec:vwo} we defined the Variance Weighted Overlap ($P_{\rm VWO}$) as a measure of bias, which depends on the correlation between biases in different regions. These correlations describe what we call the \textit{shape of the knowledge matrix}. This is illustrated in two dimensions by the ellipticity and orientation of the dark hashed ellipse in Figure~\ref{fig:heuristic}. We plot the effect of having differently shaped knowledge matrices in Figure~\ref{fig:requirements_plot_1}. Here the $P_{\rm VWO}$ as a function of the variance is plotted as a function of the bias normalised against the statistical variance: $\langle b^2 \rangle / \langle \sigma ^2 \rangle$.\footnote{These are computed by rescaling each element of $C$ by a constant factor so that ${\rm Tr}(C) = d$ where $d$ is the number of dimensions in the PCA-basis so that $\langle \sigma ^ 2 \rangle = 1$. Then $K$ is normalised so that ${\rm Tr}(K) = d b^2 /4$ in the basis where $K$ is diagonal. In 1D, this reduces to the normalisation convention used in M12.}
\par Four cases are considered. The first two are extremes, while the second two are more realistic:
\begin{itemize}
\item{$K$ has the same shape as the PC covariance $C$, appropriately normalised. $P_{VWO} > 0.95$ requires $\langle b^2 \rangle / \langle \sigma ^2 \rangle < 0.15$.}
\item{ $K$ has the same shape as the inverse PC covariance, $C^{-1}$. $P_{VWO} > 0.95$ requires $\langle b^2 \rangle / \langle \sigma ^2 \rangle < 0.005$ . } 
\item{Lensing PCs are completely independent from the techniques used to generate power spectra, so it is reasonable to assume that the bias on individual PCs are uncorrelated in which case $K$ is proportional to the identity. $P_{VWO} > 0.95$ requires $\langle b^2 \rangle / \langle \sigma ^2 \rangle < 0.02$.}
\item {In the limit of linear growth all $k$-modes are independent and the power spectrum grows according to a growth factor. It is then reasonable to assume that the bias are uncorrelated for different $k$-bins, but maximally correlated at different redshifts for fixed $k$. $K$ is computed in $k$-$z$ and then rotated to PCA-space. \textit{We refer to this as the fiducial shape and use this shape in all that follows}. $P_{VWO} > 0.95$ requires $\langle b^2 \rangle / \langle \sigma ^2 \rangle < 0.01$.}
\end{itemize}
Even in the last two most realistic cases the degeneracy between modelling errors in different regions changes the requirements on the modelling bias by a factor of two.

\subsection{Simulation Requirements} \label{sim section}
\par We now put constraints on the magnitude of the bias, $|b|$, for a Euclid-like survey, assuming $K$ has the fiducial shape (modelling assumptions are listed in Appendix \ref{sec:model choice}). In practice a knowledge matrix should be estimated from a real simulation, but in this section we examine a few test cases which are motivated above. For applications to real simulations our code, {\tt RequiSim}, is made public.
\par Figure~\ref{fig:requirements_plot_3} shows the $P_{\rm VWO}$ as a function of $|b|$, assumed to be constant everywhere, for different $\ell$-mode cuts. Ensuring that $P_{\rm VWO} > 0.95$ requires $|b| < 1 \% (3 \%)$ for $\ell_{cut} = 3000 (500)$. In practice, there are other contributions to the error budget beyond modelling the power spectrum, so the requirement on $P_{VWO}$ should be made more stringent.
\par Finally in Figure \ref{fig:requirements_plot_4}, we consider the more realistic case in which the bias varies across some transition wave-number: $k = k_b$. We assume $10\%$ bias above $k_b$ and $1 \%$ below. 
If the power spectrum can be modelled to within $1 \%$ down to scales $k_b = 7 h \text{ Mpc} ^ {-1}$, then $P_{VWO} > 0.95$ for all $\ell$-cuts. However, as $k_b$ is decreased corresponding to lowering the $k$-mode above which one feels confident in the accuracy of a simulation, $P_{\rm VWO}$ start to fall off. This is particularly noticeable if we use a large $\ell_{max}$. In fact if there is a $10 \%$ bias above $k_b = 1.1 h \text{ Mpc} ^ {-1}$ and only $1 \%$ below, then we would need to take an $\ell_{cut} < 1000$ to get an unbiased result.
\par Meanwhile, since the lensing signal is extremely insensitive to large $k$-modes, models of the power spectrum at large $k$ can be extremely inaccurate without biasing results. For example, using an $\ell_{cut} = 3000$ assuming a bias of $50 \%$ above $k_b = 7 h \text{ Mpc} ^ {-1}$ and $0.5 \%$ below still yields $P_{VWO} > 0.95$.
\par With these constraints, accurately modelling baryonic effects that affect large scales below $k = 1   h \text{ Mpc} ^ {-1}$, like AGN feedback \cite{van2011effects}, must take priority. Accurately modelling small scales effects, like radiative gas cooling \cite{van2011effects} which only becomes important beyond $k = 5   h \text{ Mpc} ^ {-1}$, is less important for cosmic shear studies.

    \section{Conclusion}
We have compared the sensitivity of tomographic with different binning strategies and 3D cosmic shear to the power spectrum and lensing kernel, independently from any assumptions about the underlying cosmological model. We draw the following conclusions:
\begin{itemize}
\item{While 3D cosmic shear captures the full 3D information content, it is slow and difficult to compute.}
\item{While equal number of galaxies per bin tomography captures the majority of the information with very few bins and is computational straightforward, this technique loses information at high-redshifts. This is where the cross-correlations with CMB lensing will be strongest.}
\item{Equally spaced redshift bin tomography does not capture as much information as equal number of galaxies per bin tomography with very few bins, however it incurs little computational cost to increase the number of bins to capture all the 3D information. We estimate that $99\% (99.9 \%)$ of the information from both the lensing kernel and the power spectrum will be captured with $50(90)$ bins.}
\end{itemize}
 \par Nevertheless the large covariance produced with a large number of bins may pose a challenge for a full likelihood analysis. Using our generalised formalism it should be possible to construct a weight function that retains the speed advantage of tomography while capturing the majority of information in just a few modes. This is left to a future work.
Meanwhile the majority of the structure growth information is extracted from the power spectrum in the region $k \in [1 h \text{ Mpc} ^{-1},  7 h \text{ Mpc} ^{-1}]$ and $z \in [0.1, 1.0]$.
\par Sensitivity to such high-$k$ modes poses a problem. Non-linear and baryonic physics which are hard to model become important at these scales. We have investigated how bias from incorrectly modelling these scales propagates to bias in the signal. 
\par Generalising the analysis in~\cite{massey2012origins} to higher dimensions, we have shown that requirements depend not only on the magnitude of the bias but where they occur in $k$-$z$ space and on the correlation between biases at different scales and redshifts.
\par Assuming that the biases are maximally correlated in redshift along fixed $k$, and uncorrelated for different $k$-modes, as they would be in the limit of linear growth and that the bias is the same everywhere, we find the power spectrum must be modelled to at least $1 \%$ accuracy for $k\leq 7 h \text{ Mpc} ^{-1}$. There are also other sources of bias so the power spectrum should be modelled more accurately than this so that it does not subsume all of its error budget allocation. This will depend on the extent to which other systematics are brought under control.
\par Unless correlations between errors in different regions of the power spectrum are extremely anti-correlated with the lensing PCs, then current simulations are not at the stage where they can be used without taking an $\ell$-cut. The stated accuracy of {\tt HALOFIT} \cite{Halofit} is $5 \%$ for $k\leq 1 h \text{ Mpc} ^{-1}$ and $10 \%$ for $k\leq 10 h \text{ Mpc} ^{-1}$. Meanwhile {\tt COSMIC EMU} \cite{heitmann2013coyote} report $4 \%$ accuracy for $k \in  [0.1 h \text{ Mpc} ^{-1}, 10 h \text{ Mpc} ^{-1}]$ and {\tt HMCode} \cite{mead2015accurate} report $5 \%$ accuracy for $k \in  [0.1 h \text{ Mpc} ^{-1}, 10 h \text{ Mpc} ^{-1}]$.   
\par Our assumptions are likely over-simplistic, so we the provide public code, {\tt RequiSim}, to compute the bias on the lensing signal from inaccurate power spectrum models produced by any simulation.
\par Although we have not computed the bias on the lensing signal for existing power spectrum codes, we can provide a qualitative road map forward for simulators. Since the power spectrum is largely insensitive to scales $k > 7 h \text{ Mpc} ^ {-1}$, simulators should focus on accurately modelling scales of $k < 7 h \text{ Mpc} ^ {-1}$ first.

\section*{Acknowledgements}

    The authors are grateful for constructive conversations with Alan Heavens, Alessio Spurio Mancini, Robert Reischke, Bj\"orn Malte Sch\"afer and Alexander Mead.
   We thank the {\tt Cosmosis} team for making their code publicly available. 
    PT is supported by the UK Science and Technology Facilities Council.
    TK is supported by the Royal Society.
    The authors acknowledge the support of the Leverhume Trust.

\appendix

\section{Motivation for the Bessel Weight} \label{sec:A}
\par There are in general three reasons to write a signal in spherical-Bessel space:
\begin{enumerate}
\item Spherical-Bessel functions follow an orthogonality relation~\cite{castro}.
\item Spherical-Bessel functions and spherical-harmonics are eigenfunctions of the Laplace operator in spherical coordinates. This ostensibly comes from the Laplacian used to relate the lensing potential and hence the shear in terms of the density field through the power spectrum.
\item If the observed signal traces the cosmological density field directly, e.g. as is the case in galaxy clustering, then at a redshift $z$, the projected power spectrum of the signal in a spherical-Bessel representation, $C_{\ell} \left(k ; z \right)$, is related to the matter power spectrum, $P \left( k ; z \right)$, by an equality \cite{castro}:   
\begin{equation} 
C_{\ell} \left(k ; z \right) = P \left( k ; z \right).
\end{equation}
This implies cutting out high $k$-modes from the observed projected spectrum should cleanly remove sensitivity to small poorly understood, high-$k$ mode, scales in the matter power spectrum~\cite{lanusse20153d}.
\end{enumerate}
\par The first consideration is a valid reason to use a spherical-Bessel basis set for weak lensing as it ensures that the shot noise (see equation \ref{eq:Noise}) is uncorrelated, and does not become too large. However \emph{any} orthogonal set of function will suffice in equation \ref{sB} and the Bessel functions are needlessly expensive to compute compared to other choices. 
\par The second consideration is not relevant here because only the Newtonian potential must be expressed in this basis to relate it to the cosmological density field, through a Poisson equation; this is where the Bessel functions in equation~\ref{eq:U} originate~\cite{castro}.
\par Meanwhile the lensing power spectrum, $C_l$, traces all matter power \textit{below} a certain redshift, weighted by a lensing kernel i.e.~the power spectrum is always enclosed within an integral over the line-of-sight. Hence, there is no reason that taking an $\eta$-cut should preferentially remove sensitivity to small scales in the matter power spectrum. However confusion can arise by labelling both the lensing spectrum and the power spectrum wave-number with $k$, and equating the two. We discuss directly removing sensitivity to small scales in further in Appendix~\ref{sec:H}.

    \section{Fisher Matrix Formalism} \label{sec:B}
\par Before conducting an experiment, the Fisher matrix can be used to estimate constraints and predict degeneracies for a set of parameters, $\{ \theta_i \}$~\cite{tegmark1997karhunen}. We use it to estimate constraints on different regions of the matter power spectrum and lensing kernel, and predict correlations between them. 
\par  Provided the likelihood is Gaussian, the Fisher matrix for cosmic shear is: 
\begin{equation} \label{eqn:fish}
F_{ij} = \sum_{\ell} \frac{2\ell+1}{2} Tr \left[ C_{\ell}^{-1} C_{{\ell},i} C_{\ell}^{-1} C_{{\ell},j} \right],
\end{equation}
where $C_{{\ell},i}$ is the derivative of the lensing spectrum with respect to parameter $\theta_i$.
\par Normally the sensitivity to the original parameters is given by the covariance matrix, $C$, found by inverting the Fisher matrix. However, in our analysis we produce many large, ill-conditioned and nearly singular Fisher matrices, so inversion introduces too many numerical artefacts. Instead we use the Cramer-Rao Bound. Defining $\sigma_i$ as the conditional error on $\theta_i$, the Cramer-Rao Bound is:
\begin{equation} \label{eq:inverse error}
\frac{1}{\sigma_i} \leq \sqrt{F_{ii}},
\end{equation}
assuming all other parameters are known \cite{tegmark1997karhunen}.
This measure of uncertainty does not account for the correlations between parameters. An alternative measure, which does, is defined in Section~\ref{sec:PCA}.

\section{Principal Component Analysis (PCA)} \label{sec:PCA}
For any experiment we can choose a set of parameters and estimate their covariance. A Principal Component Analysis (PCA) finds a smaller set of independent parameters that capture the majority of the information. Informally these can be thought of as the parameters that are actually being measured, or that the data is in fact sensitive to.
\par For a set of $N$ parameters, $\{ \theta_i \}$, the covariance matrix, 
$C\equiv F^{-1}$, encodes parameter degeneracies. Since it is symmetric it can be rotated into an eigenbasis where there are no degeneracies:
\begin{equation} \label{eq:cov}
C = P ^{T} D P,
\end{equation}
where $D$ is a diagonal matrix of $N$ non-zero eigenvalues, $P$ is a matrix formed of real eigenvectors of $C$, and  $P^T$ is the transpose of $P$. 
\par These new parameters, $\{ \zeta_i \}$, are related to the old parameters by:
\begin{equation} \label{eq:old params}
\zeta_i = \sum _j \left(v_i \right) _j \theta_j,
\end{equation}
where $v_i$ is the $i$th eigenvector of $C$, and the $i$th row of the matrix $P$.
When we apply this formalism to the power spectrum and lensing kernel, the set $\{ \zeta_i \}$ will correspond to the amplitudes of a set of step functions, $\{ f_i \}$, which can be formed from $\{ v_i \}$. We refer to these functions as \textit{components} and the value of the $j$th element of $v_i$ denotes the height of cell $j$ in component $i$. For power spectrum and lensing kernel PCs the cells will define regions in $k$-$z$ space and $z$ space, respectively (see next section for more details).
\par The components with the smallest eigenvalues are the most tightly constrained and hence they contain the most information. Arranging the components according to ascending order in the corresponding eigenvalues, $\lambda_i$, also the diagonal components of, $D$, we define the fractional information content of the first $m$ eigenvalues as:
\begin{equation} \label{eq:info}
I \left( m \right) = \frac{\sum \limits_{{i<m}} \lambda_i} {I_{\text{tot}}},
\end{equation}
where $I_{\text{tot}} \equiv \sum_i  \lambda_i$. The first few components, that contain the majority of the information, are called the \textit{principal components} (PCs). Meanwhile the \textit{total information content}, $I_{\text{tot}}$, is also occasionally referred to as a the \textit{Figure of Merit} (FoM). This measure will be useful to compare the total constraining power of 3D and tomographic cosmic shear.
\par To avoid inverting the Fisher matrix which may be ill-condtioned, we will calculate the PCs directly. Inverting, $C$ in equation~\ref{eq:cov}, we find:
\begin{equation}
F = P^T D^{-1} P,
\end{equation} 
and the PCs correspond to the largest eigenvalues along the diagonal of $D^{-1}$. 
\par The total sensitivity to different regions can also be found without inverting $F$ by taking a weighted sum of the components, $f_i$, in terms of the information content of each. This is given by the \textit{Variance Weighted Sum} defined as:
\begin{equation}
S_{vw} = \sum _i \lambda_i | f_i|.
\end{equation}
The absolute value is taken because individual components have positive and negative values. Since $S_{vw}$ is computed in PCA space it naturally takes into account correlations between components unlike the Cramer-Rao Bound.

\section{Power Spectrum and Lensing Kernel Principal Components} \label{sec:D}
\par To determine how 3D and tomographic cosmic shear are sensitive to the matter power spectrum and the lensing kernel, we perform a PCA, closely following the procedure in \cite{pca}. Our analysis asses the sensitivity to the growth of structure and background evolution independently from any assumption of the underlying cosmological model.
\par To find the power spectrum PCs, we divide the power spectrum, $P \left( k, z \right)$, into logarithmically and linearly spaced grid cells in $k$ and $z$, respectively. Inside each grid cell, $i$, we compute the fractional amplitude change in the power:
\begin{equation}
	P_i \left(k,z, \mathcal{A} \right)  \equiv
    \begin{cases}
      \left( 1 + \mathcal{A}\right) P \left(k,z \right) & \text{if $(k,z)$ in cell $i$  }\\
      P \left(k,z \right) & \text{otherwise},\\
    \end{cases}
\end{equation}
where $\mathcal{A}$ is a fixed small amplitude change.  Defining each of these transformations as a parameter, $\theta_i$, we compute a two sided derivative:
\begin{equation}
\frac{\partial P}{\partial \theta_i } = \frac{P_i \left(k,z, \mathcal{A} \right) - P_i \left(k,z, -\mathcal{A} \right)}{2 \mathcal{A}}.
\end{equation} 
\par From these we compute the Fisher matrix, and hence the PCs. In \cite{pca}, the authors computed the PCs on a low resolution matter power spectrum grid, before smoothly interpolating to higher resolution. Therefore it is unclear how much of the structure seen in their PCs is due to interpolation errors. Our method avoids this issue since the matter power spectrum is perturbed only after interpolation. Interpolation errors can thus be seen as a small change to the fiducial power spectrum.
\par We find the lensing kernel PCs by dividing the co-moving distance into equally spaced redshift slices and making the perturbation:
\begin{equation}
	 r_i \left(z, \mathcal{A} \right) \equiv
    \begin{cases}
      \left( 1 + \mathcal{A}\right) r \left(z \right) & \text{if $z$ in slice $i$ }\\
     r \left(z \right) & \text{otherwise},\\
    \end{cases}
\end{equation}
Hence the perturbed lensing kernel is:
\begin{equation}
\left( F_k \right)_i  \left(  \mathcal{A} \right) = \frac{r_i\left(z, \mathcal{A} \right)  - r'_i\left(z, \mathcal{A} \right) }{r_i\left(z, \mathcal{A} \right) r'_i\left(z, \mathcal{A} \right)  }.
\end{equation}
Again treating each perturbation as a separate parameter, $\theta_i$, we define the two sided derivative as:
\begin{equation}
\frac{\partial F_k }{\partial \theta_i} = \frac{  \left( F_k \right)_i  \left(\mathcal{A} \right) -  \left( F_k \right)_i  \left(\mathcal{-A} \right)}{2 \mathcal{A}},
\end{equation}
and compute the Fisher matrix as before.

\par In theory there are correlations between power spectrum PCs and lensing kernel PCs inside a much larger Fisher. However perturbations to the power spectrum have a very different effect on the lensing signal to perturbations to the lensing kernel so we assume the two types of PCs are uncorrelated.

    \section{Modelling Choices } \label{sec:model choice}
We assume a Gaussian distribution for the photometric redshift error given by:
\begin{equation} \label{eq:photo error}
p \left( z | z_p \right) \equiv \frac{1}{2 \pi \sigma_z \left(z_p \right)} e ^{- \frac{ \left( z -c_{cal} z_p + z_{bias} \right) ^2  } {2 \sigma_{z_p}} },
\end{equation}
with $c_{cal} = 1$, $z_{bias} = 0$ and $\sigma_{z_p} = A \left(  1 + z_p \right)$ with $ A = 0.05$ \cite{ilbert2006accurate} and
\begin{equation} \label{eq:n(z)}
n \left( z_p \right) \propto \frac{a_1}{c_1} e ^ {- \frac{ \left( z-0.7 \right) ^2 }{b_1^2} } + e ^ {- \frac{ \left( z-1.2 \right) ^2 }{d_1^2} } ,
\end{equation}
with $\left(a_1/c_1,b_1 ,d_1 \right) =\left( 1.5 / 0.2, 0.32, 0.46 \right)$ \cite{van2013cfhtlens}. We assume a 15,000 degree survey with 30 galaxies per $\text{arcmin} ^2$.
\par We use a fiducial LCDM cosmology with $\left( \Omega_m, \Omega_k, \Omega_b,  h_0, n_s, A_s,  \tau  \right) = \left(0.315,\text{ }  0.0,\text{ } 0.04,\text{ }  0.67,\text{ } 0.96,\text{ } 2.1 \times 10 ^9,\text{ } 0.08 \right)$ throughout.  
The power spectrum is generated using {\tt CAMB} \cite{camb} and the non-linear part is generated using {\tt HALOFIT} \cite{Halofit}, produced as part of the {\tt Cosmosis} \cite{cosmosis} pipeline, each run with the default setting given in the \textit{demo1} tutorial in {\tt Cosmosis}.

 \section{Appendix: Convergence Checks} \label{sec:convergence checks}
 
    \begin{figure}
   \centering
    \vspace{2mm}
    \includegraphics[width=85mm]{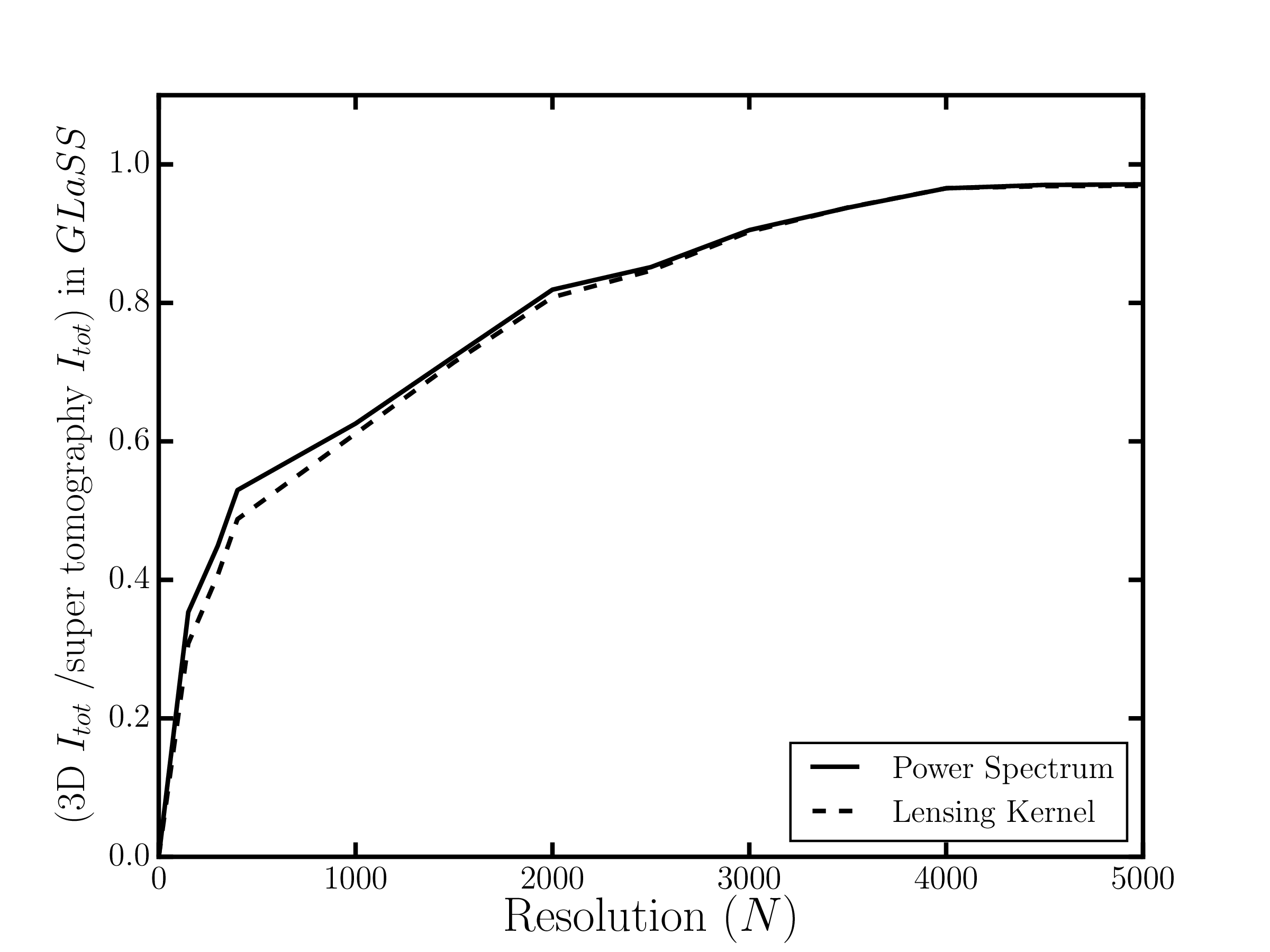}
    \caption{Convergence of 3D cosmic shear total information content, $I_{tot}$, (see equation~\ref{eq:info}) relative to super tomography as a function of the resolution of the computation grid. Both lensing kernel and power spectrum $I_{tot}$ converges to within $3 \%$ at a resolution of $N = 5000$ and $20 \%$ at $N = 2000$. $I_{tot}$s were computed on a coarse PCA grid sampled very sparsely in $\ell$. Due to memory constraints $N = 2000$ throughout the rest of the paper.}
    \label{fig:convergence}
    \end{figure}
 
To reduce computation time, we compute the Fisher matrix sampling sparsely in $\ell$, taking:
\begin{equation}
F_{ij} = \sum_{\ell \in L} \Delta \ell \frac{2\ell+1}{2} Tr \left[ C_\ell^{-1} C_{\ell,i} C_\ell^{-1} C_{l,j} \right],
\end{equation}
where  $\ell$ is sampled at $[2, 12, 25, 50, 75]$ below $100$, then at intervals of $50$ to $\ell= 2000$ and finally intervals of $100$ to $\ell = 3000$. This cuts the computation time by nearly an order of magnitude. For $10$-bin tomography this leads to a $<7 \%$ error inside the variance weighted sum in the largest power spectrum PCA bin and $<0.1 \%$ average error across all bins, compared to the Fisher where every $\ell$-mode is sampled. 
\par The lensing power spectrum is computed on an $N \times N$ grid logarithmically spaced in $k$ and linearly space in $z$. For our analysis tomography and super tomography are fully converged for $N = 400$, which we have used throughout. As 3D cosmic shear converges slowly, we test the convergence of 3D cosmic shear on a coarse PCA grid: two-by-two and two-by-one for the power spectrum and lensing kernel respectively. The convergence of the information content as a function of $N$, relative to super tomography, is shown in Figure~\ref{fig:convergence}. 

\section{Comparison with Other Work} \label{sec:comparison}
Tomographic and 3D cosmic shear were recently compared in \cite{spuriomancini} (hereafter SM18) which reports a decrease in the error on some modified gravity parameters of $20 - 30 \%$ for 3D cosmic shear compared to $6$-bin tomography with an equal number of galaxies per bin. Meanwhile we only find a $15 \%$ and $2\%$ increase in the total information, $I_{tot}$, for the lensing kernel and power spectrum respectively when going from this regime to super tomography. 
\par The slightly smaller gains in our analysis, are expected due to two differences in modelling assumptions. SM18 used $\ell_{max} = 1000$ while we used $\ell_{max} = 3000$. The higher $\ell$-cut used in our analysis means we are relatively more sensitive to lower redshifts below $z=0.5$ (see Figure \ref{fig:pca_l_cuts}). However, tomography with an equal number of galaxies per bin primarily loses information at higher redshifts, beyond $z=1$ (see Figure \ref{fig:tomo_kernel_more_bins_fig}). SM18 used a linear power spectrum, while we used a non-linear which relatively boosts our sensitivity to high-$k$ modes in the power spectrum. Again these modes are primarily probed at low-$z$ (see Figure \ref{fig:tomo_fig_high_res}) where tomography with an equal number of galaxies per bin loses information.

 \section{Directly removing Sensitivity to Small Scale Power} \label{sec:H}
\par As small scales modelling error introduce bias. Ideally it would be possible to remove sensitivity to these modes above some $k_{cut}$. We split the matter power spectrum into two parts: $P^{k >k_{cut}}$ and $P^{k <k_{cut}}$, where the former contains only power above the cut and the later power below. The resulting lensing spectra: $C_\ell^{k >k_{cut}}$ and $C_\ell^{k<k_{cut}}$, were calculated. The spectra have power at nearly identical modes making it difficult to reduce the sensitivity to small scales without also losing sensitivity to the signal. 

 \section{RequiSim} \label{sec:I}
{\tt RequiSim} is available for download from: \url{https://github.com/astro-informatics/RequiSim}. Using pre-computed PCs and a user-provided knowledge matrix, {\tt RequiSim} computes the $P_{VWO}$, for a Euclid-like survey. PCs for other surveys can be computed on request. 

\bibliographystyle{apsrev4-1.bst}
\bibliography{bibtex.bib}

\end{document}